\documentclass[useAMS,usenatbib]{mn2e}

\usepackage{latexsym}
\usepackage{afterpage}
\usepackage{longtable}
\usepackage[dvips]{graphicx}
\usepackage{natbib}
\usepackage{psfrag}
\usepackage{rotating}
\usepackage{amssymb}
\usepackage{pdflscape}

\def\Hzero{$\rm H_0$}
\def\OM{$\rm \Omega_M$}
\def\OL{$\rm \Omega_{\Lambda}$}

\def\SoneG{\rm S_{\rm 1.4GHz}}

\def\logP{\rm logP_{1.4GHz}}
\def\uniP{\rm \, W\,Hz^{-1}sr^{-1}}

\def\AJ{{\it Astron. J.}}
\def\ApJ{{\it Astroph. J.}}
\def\ApJS{{\it Astroph. J. Sup.}}
\def\MNRAS{{\it Mon. Not. R. Astr. Soc.}}
\def\AandA{{\it Astron. Astrophys.}}

\def\ASPC{{\it Astr. Soc. Pac. Conf. Ser.}}

\title[Morphology, accretion modes and environment in AGN]{The relation between morphology, accretion modes and environmental factors in local radio AGN}

\author[Melanie A. Gendre, P. N. Best, J. V. Wall, \& L. M. Ker]{M. A. Gendre$^{1}$\thanks{E-mail:
    mgendre@jb.man.ac.uk}, P. N. Best$^{2}$, J. V. Wall$^{3}$ and L. M. Ker$^{2}$\\ 
$^{1}$Jodrell Bank Center for Astrophysics, The University of Manchester, Oxford Rd, Manchester M13 9PL, United Kingdom\\
$^{2}$Institute for Astronomy, Royal Observatory, Blackford Hill,
Edinburgh EH9 3HJ, United Kingdom\\
$^{3}$Department of Physics and Astronomy, The University of British
  Colombia, 6224 Agricultural Rd, Vancouver, BC, V6T 1Z1,\\ 
\phantom{$^{2}$}Canada}

\begin{document}

\date{Accepted . Received ; in original form }

\pagerange{\pageref{firstpage}--\pageref{lastpage}} \pubyear{}

\maketitle

\label{firstpage}

\begin{abstract}
The goal of this work is to determine the nature of the relation between morphology and accretion mode in radio galaxies, including environmental parameters.  The CoNFIG extended catalogue (improved by new K$_S$-band identifications and estimated redshifts from UKIDSS, and spectral index measurements from new GMRT observations) is used to select a sub-sample of 206 radio galaxies with z$\le$0.3 over a wide range of radio luminosity, which are morphology-classified using the Fanaroff-Riley (FR) classification of extended radio sources. For each galaxy, spectroscopic data are retrieved to determine the high/low excitation status of the source, related to its accretion mode. Environmental factors, such as the host galaxy luminosity and a richness factor are also computed, generally using SDSS data.  We find the following results: (1) At a given radio luminosity, the FR morphological split of sources is consistent with being the same for both accretion modes. This remains true if analysis is restricted to only rich or only poor environments. If confirmed with a larger sample, this would imply that extended radio morphology is independent of the accretion mode of the black hole, depending only on the power of the resultant jet, and its interactions with the larger-scale environment. (2) Excitation modes seem to be linked to the source environment, with high-excitation galaxies found almost exclusively in low-density environments while low-excitation galaxies occupy a wider range of densities; this result is independent of FR morphology, and is consistent with the different fuelling mechanisms expected for these excitation modes. (3) Independent of excitation mode, FRI sources are found to lie in higher density environments, on average, than FRII sources, consistent with FRI sources having their jets disrupted by a denser surrounding medium. However, there is a significant overlap in environment between the two classes, and no clear driving factor between the FRI and FRII sources is found even when combining radio luminosity, accretion mode, large-scale environment and host galaxy luminosity.
\end{abstract}

\begin{keywords}
catalogues - radio continuum: galaxies - galaxies: active - galaxies: statistics - galaxies: luminosity function
\end{keywords}

\section{Introduction}

Extended radio-loud AGN can be classified according to their morphology, following the Fanaroff-Riley (FR) scheme \citep{FR74}, in which FRI objects have the highest surface brightness along the jets near the core, while FRII sources show the highest surface brightness at the lobe extremities, as well as more collimated jets. The division between FRI and FRII is however somewhat ambiguous, with the existence of hybrid sources showing jets FRI-like on one side and FRII-like on the other \citep{Capetti95,Gop00}.

The FR dichotomy is based purely on the appearance of the radio objects, and the mechanisms differentiating the two populations are still unknown. Two main streams of models have been postulated to explain these differences in morphology.  Extrinsic models, purely based on the interaction of the jet with the source environment, were proposed based on environmental differences found between FRI and FRII sources \citep[e.g.][]{Prest88}, and on their apparently distinct host galaxies \citep{Owen94}. The hypothesis is that inter-galactic medium (IGM) density is the differentiating factor, where jets of sources in higher/lower density mediums experience a higher/lower degree of resistance, yielding sources with FRI/FRII structures respectively. Intrinsic models, on the other hand, were postulated based on fundamental differences seen between FRI and FRII sources, such as their emission line properties \citep{Zirb95}. These models suggested that the dichotomy arises from differences in the properties of the central black hole \citep[e.g.][]{Baum95,Ghis01}. In these scenarios, jets produced by low accretion-flow rate which are generally weak, mostly display FRI-type structure, whereas galaxies with higher accretion flow rates give rise to stronger, mainly FRII-type jets.\\

More recently, these different accretion rates have been associated with the excitation mode of the narrow line region gas in the host galaxy. In low-excitation galaxies (LEG), also known as `radio-mode' or `hot-mode' accretors, the accretion onto the black hole is radiatively inefficient but does produce highly energetic radio jets via the emission of kinetic energy through the radio jets \citep{Merl07}.  High-excitation galaxies (HEG), also known as `quasar-mode' or `cold-mode' accretors, are linked to radiatively-efficient accretion disks \citep{Shak73} and are often identified with star-formation activity in the host galaxies \citep{Kauff03}. Several recent studies \citep{Hard07,Kauff08,Baldi08} suggest that HEGs have undergone a recent merger that triggered star formation, driving cold gas towards the central engine, powering the AGN (cold gas accretion). LEGs have had no such recent merger and show no evidence of recent star formation, and are believed to be fuelled by the hot inter-stellar medium (ISM), possibly as part of a feedback cycle \citep[e.g.][]{Best05}. Thus, although some other alternative explanations for the influx of cold gas in HEGs exists, such as recycled gas from dying stars \citep{Ciotti07}, mergers or interactions seem to give the most likely explanation for cold gas accretion.

\cite{Baldi08} studied nearby 3CR radio galaxies and their optical properties and found indication of recent star formation in HEGs, but not in the LEGs. In a different study, \cite{Em08} found no evidence for large-scale H{\small I} in low-luminosity sources, but significant amounts in high-luminosity sources. The `radio-mode' accretors were also shown to be fundamentally different from the `quasar-mode' accretors from X-ray and infrared observation \citep{Hard07}.  Finally, a dedicated study of HEGs and LEGs by \cite{Best12} confirmed that both population have indeed fundamentally different accretion rates (with $\rm L_{\rm HEG} \sim \rm 0.1 L_{\rm Edd}$ while $\rm L_{\rm LEG} < \rm 0.01 L_{\rm Edd}$) and host galaxy properties \citep[with LEGs being redder and larger and having more massive galaxy and black hole mass than HEGs of similar radio power; see also][]{Jan12}. 

These distinctions between HEGs and LEGs are very reminiscent of the differences between FRI and FRII sources \citep[e.g.][]{Jack99}. This is because there is a large overlap in populations between FRIs and LEGs, and between FRIIs and HEGs. However the relation is not one-to-one: small subsets of FRIs are found in HEG samples, as well as many FRIIs being associated with LEGs \cite[e.g.][]{Laing94,Willott01,Hey07,Hard07}. This implies that the FR dichotomy is not fully dependent on accretion mode.

It has long been known \citep{Long66} that the radio luminosity function undergoes luminosity-dependent evolution, where low-luminosity sources show little or no evolution while high-luminosity sources undergo positive density evolution. In an initial modelling of the space density of radio AGN, \cite{Wall97} and \cite{Jack99} assumed that this was based on a division of the radio sources into low-luminosity, non-evolving FRIs and high-luminosity, rapidly evolving FRIIs. However, more recent results have shown that, at comparable powers, FRI and FRII sources show strong similarities in evolution \citep[e.g.][]{Snellen01,Rigby08,PaperII}, whereas there are indications from the work of \cite{Best12} that the cosmic evolution of HEGs and LEGs is different at fixed radio luminosity (HEGs show evidence of strong evolution while LEGs are consistent with little evolution). This implies that LEGs and HEGs may be more appropriate as the two fundamental populations of radio-AGN \citep[see also][]{Chia00,Butti10,Herb10,Kun10}. From there, in the simplest model, the various observable radio morphology must result from external effects, such as ISM/IGM density (FRI vs. FRII) and/or jet orientation (compact vs. extended).\\

The goal of this work is to determine the nature of the relation between morphology and accretion mode in radio galaxies, including environmental parameters. It is based on the extended CoNFIG catalogue \citep{PaperI,PaperII}, which has been improved in terms of spectral index and redshift using both new GMRT radio observations and literature data (\S\ref{Config}). From there, a comparative study of the FRI/II in the Local Universe (z$\le$0.3) is performed, particularly looking into the FR morphology-accretion mode connection, including environmental parameters. For this purpose, excitation classifications, host galaxy luminosity and cluster richness measurements (from the Sloan Digital Sky Survey \citep[SDSS;][]{York00}, the SuperCosmos Sky Survey \citep[SSS;][]{Hambly01}, and the ESO Imaging Survey \citep[EIS;][]{Noni99}) were introduced to the local CoNFIG sub-sample (\S\ref{FRdicho}). The results are then discussed in \S\ref{Results}.\\

Throughout this paper, we assume a standard $\rm \Lambda$CDM cosmology with \Hzero = 70 km s$^{-1}$ Mpc$^{-1}$, \OM = 0.3 and \OL = 0.7.

\section{Improving the CoNFIG Sample}\label{Config}

\begin{figure*}
  \begin{minipage}{16.cm}
    \begin{minipage}{3.5cm}
      \centerline{
        \includegraphics[angle=270,scale=0.18]{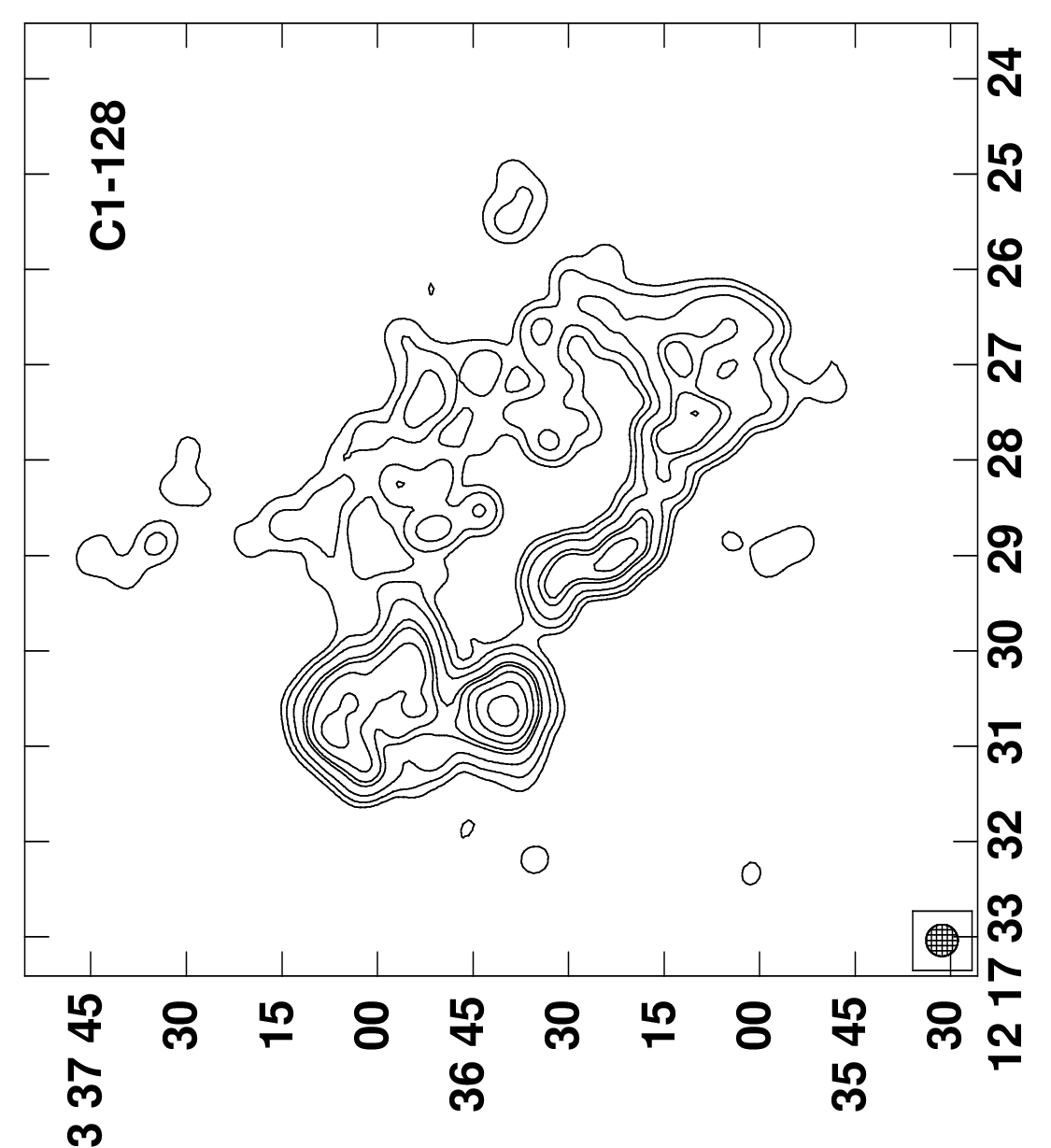}}
    \end{minipage}
    \hfill
    \begin{minipage}{3.5cm}
      \centerline{
        \includegraphics[angle=270,scale=0.18]{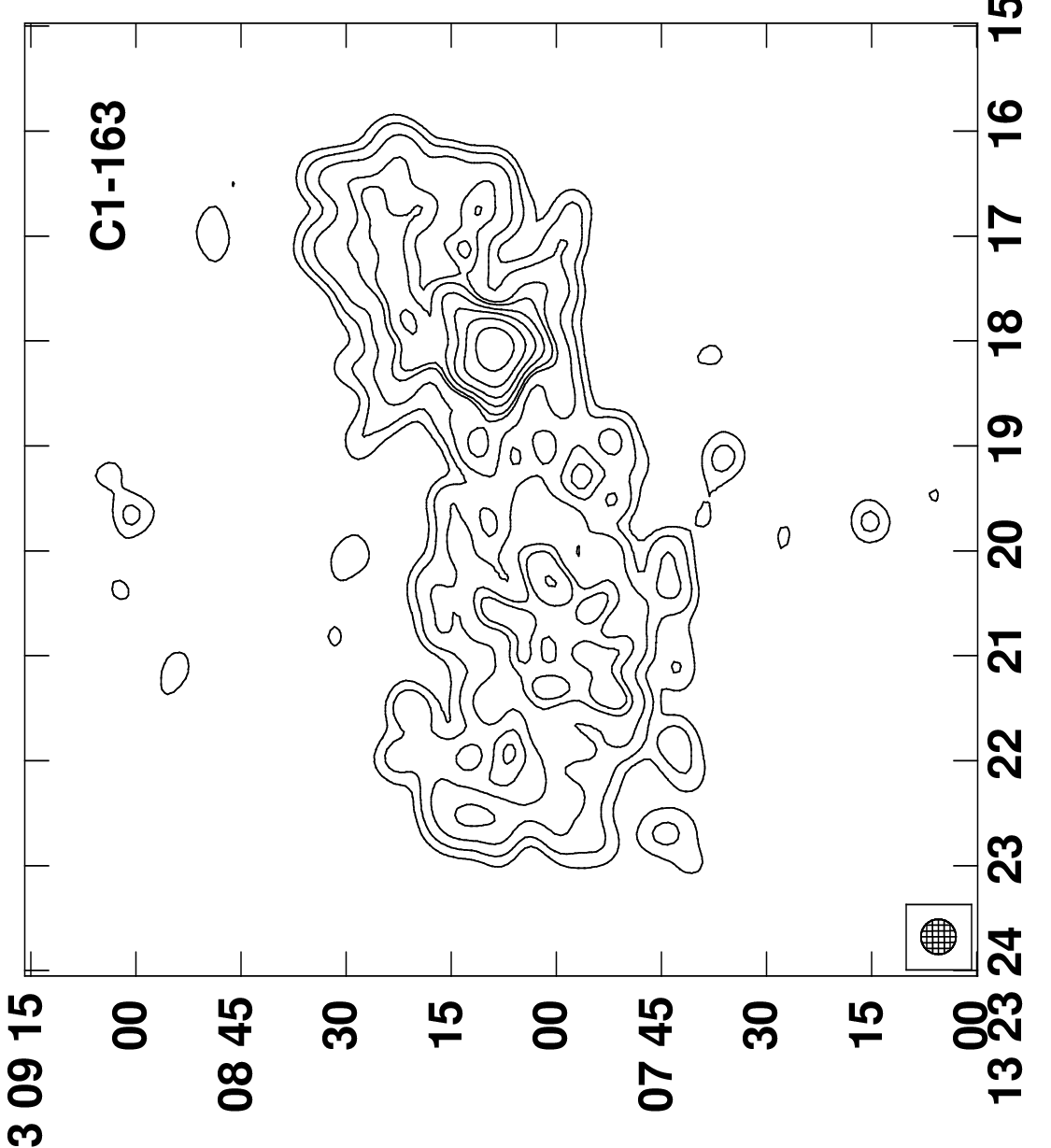}}
    \end{minipage}
    \hfill
    \begin{minipage}{3.5cm}
      \centerline{
        \includegraphics[angle=270,scale=0.18]{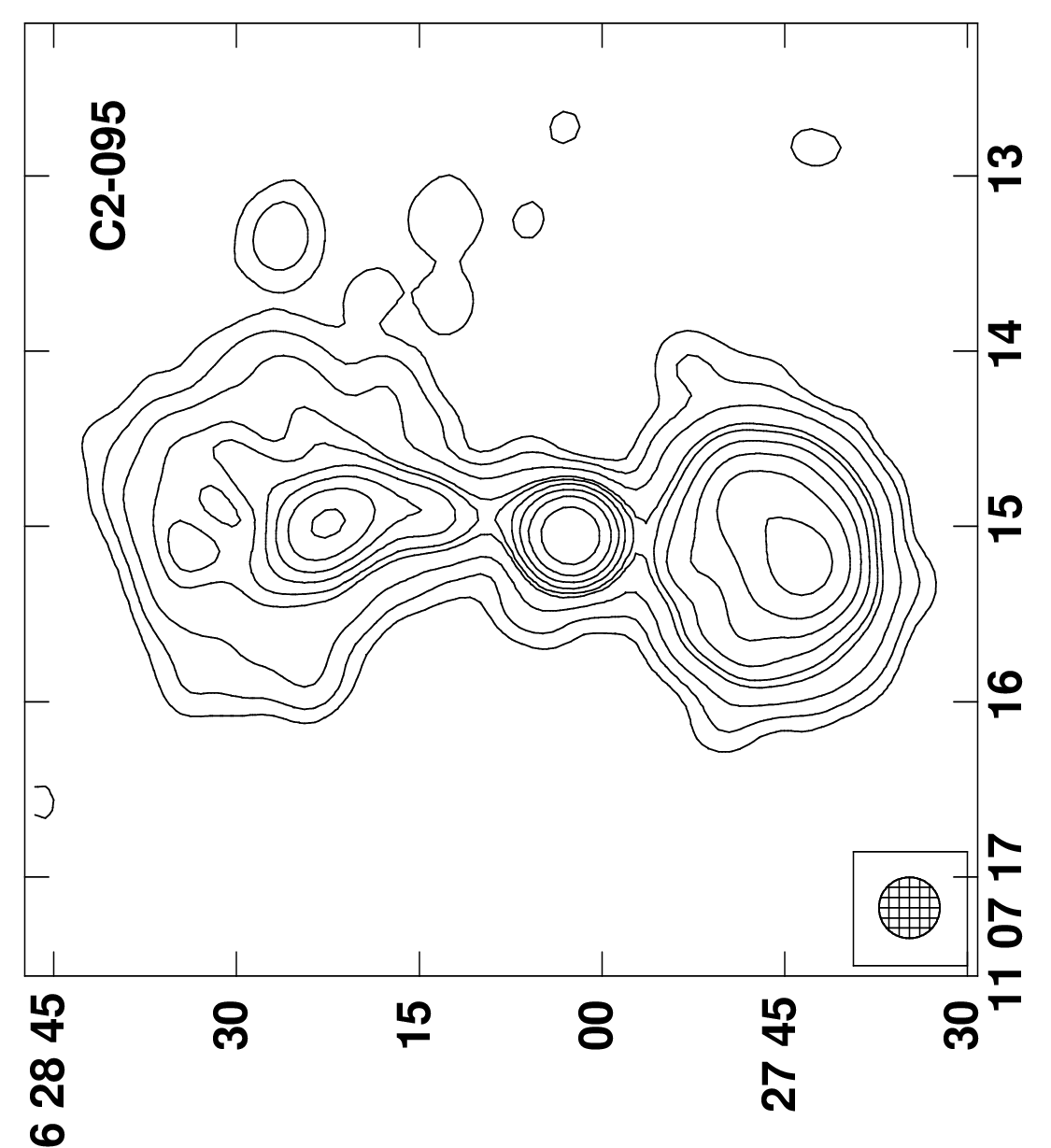}}
    \end{minipage}
    \hfill
    \begin{minipage}{3.5cm}
      \centerline{
        \includegraphics[angle=270,scale=0.18]{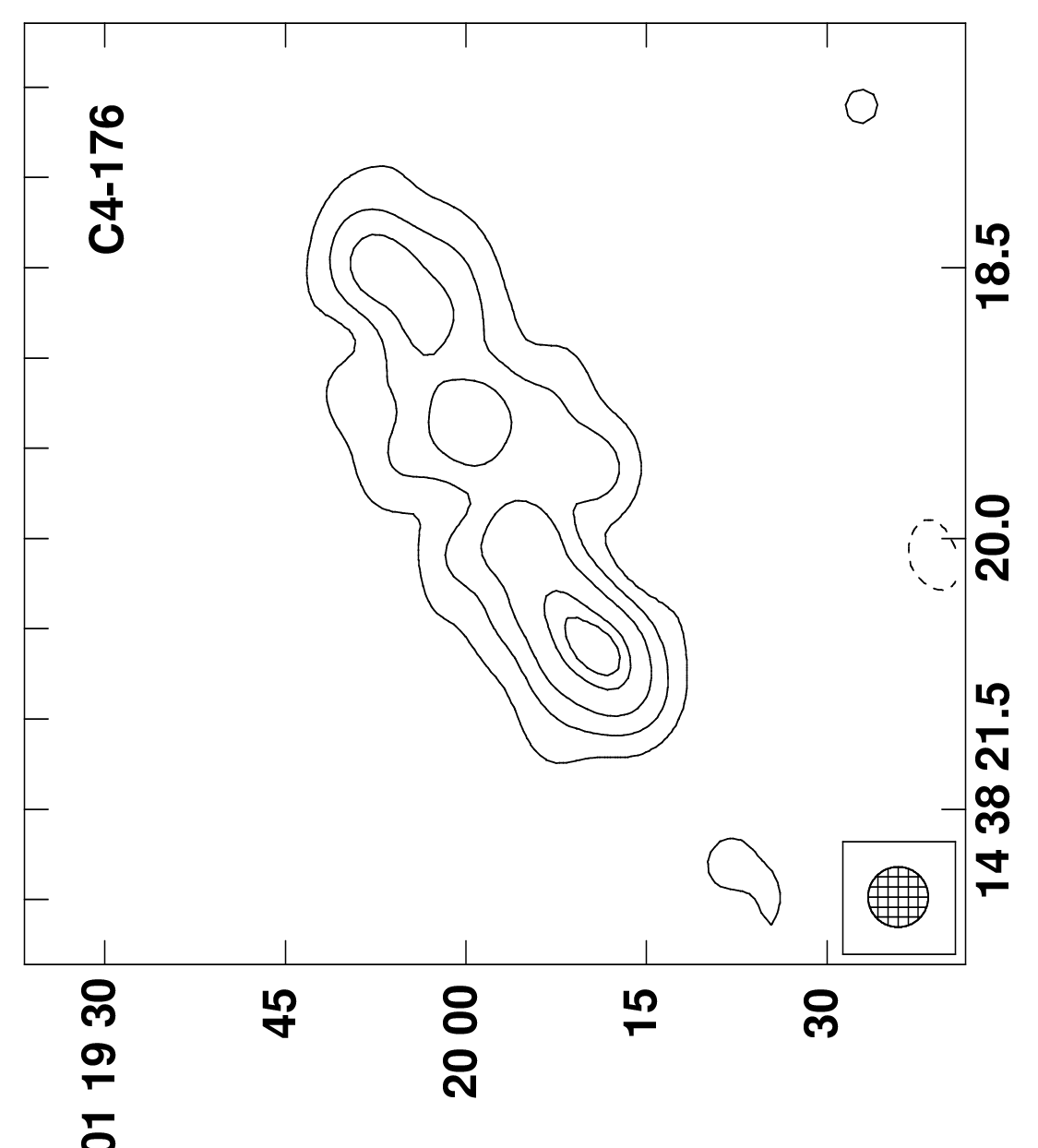}}
    \end{minipage}
  \end{minipage}
  \vspace{2mm}
  \caption[]{\label{NewMorph}GMRT 591-MHz continuum observation contour maps at -1, 1, 2, 4, 8, 10, 14, 20, 30, 50 $\times$ 1.5 mJy/bm for the sources (from left to right) 4C 04.41, 4C 03.27,4C 16.30 and 1438-0133. The reference catalogue number for each source is shown on the top right corner of each image.}
\end{figure*}
\begin{figure}
  \begin{minipage}{7.5cm}
    \centerline{
      \includegraphics[angle=270,scale=0.33]{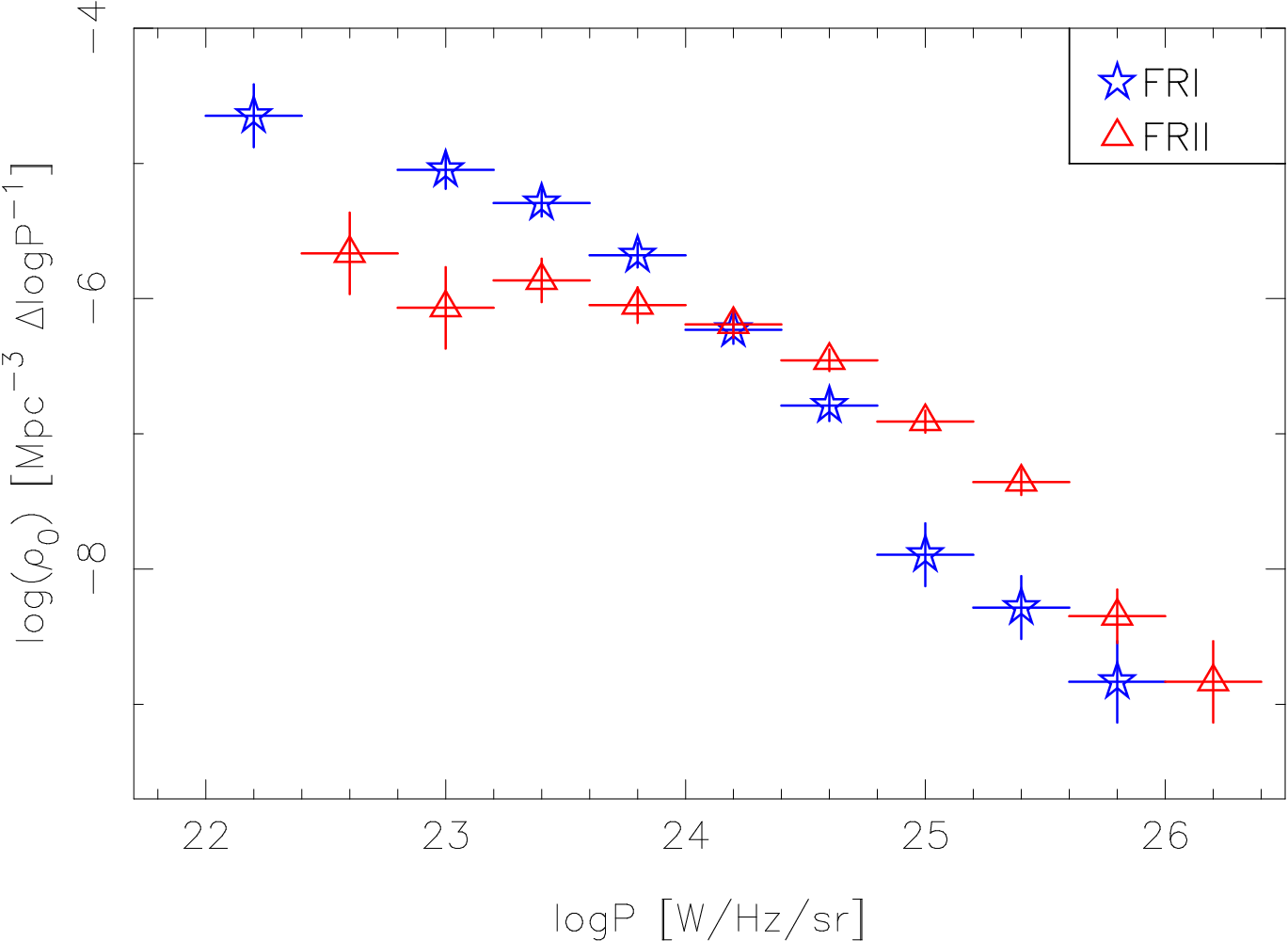}}
    \caption[]{\label{LRLF} Updated local radio luminosity function $\rho(P)$ for FRIs and FRIIs, using bin sizes of $\rm \Delta \logP =0.4$, represented by stars and triangles respectively.} 
    \end{minipage}
  \vfill
  \vspace{5mm}
  \begin{minipage}{7.5cm}
    \centerline{
      \includegraphics[angle=270,scale=0.33]{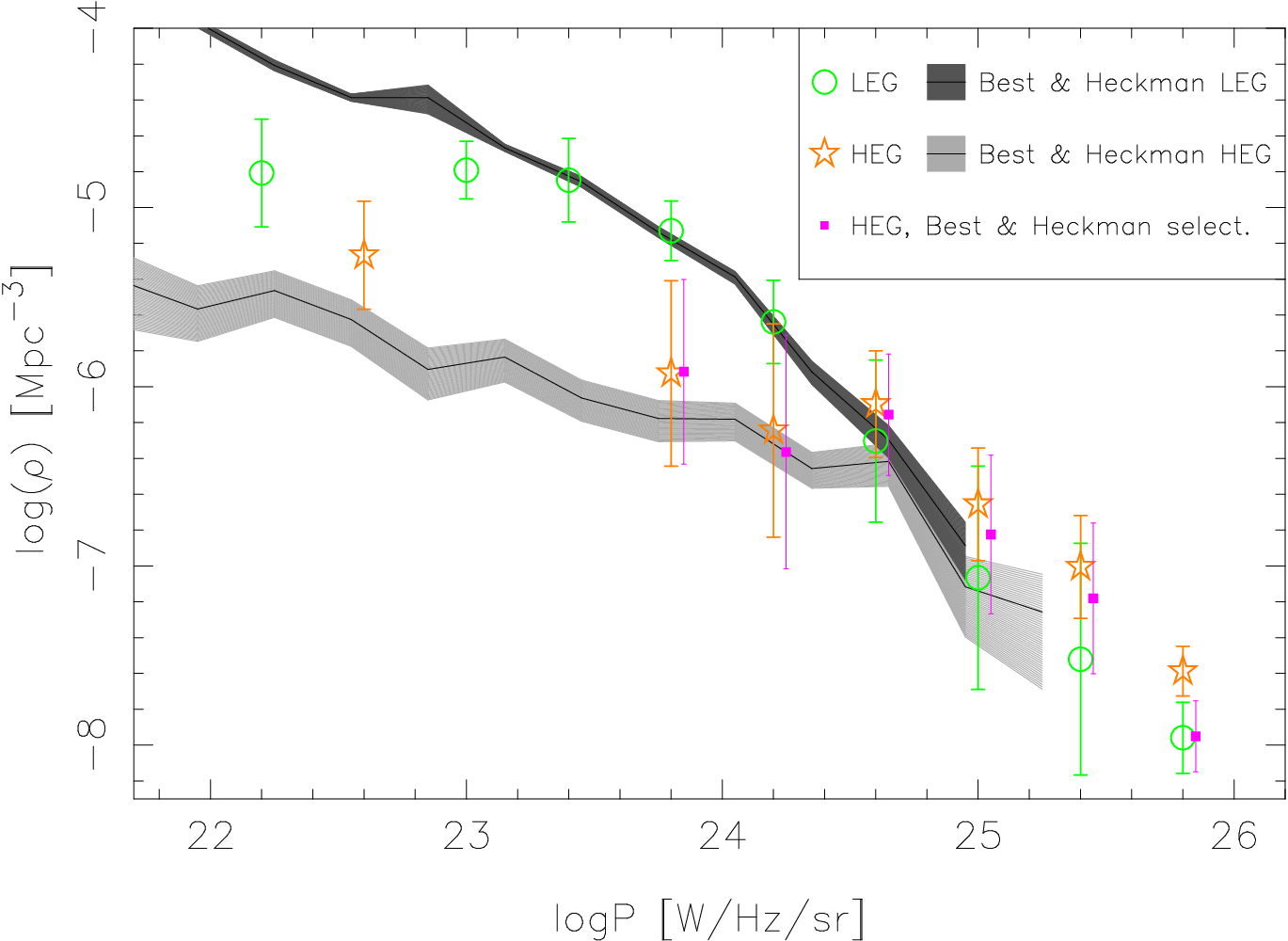}}
    \caption[]{\label{HGLRLF}Local radio luminosity function  $\rho(P)$ for HEGs (stars) and LEGs (circles) separately, using bin sizes of $\rm \Delta \logP =0.4$. The LRLFs are compared to results from \cite{Best12} (in light and dark grey for HEGs and LEGs respectively). For more accurate comparisons, the LRLF for CoNFIG HEGs with SDSS counterparts (excluding quasars) as selected by \cite{Best12} is shown in filled squares.\\}
    \end{minipage} 
\end{figure}

The extended CoNFIG catalogue \citep{PaperII} is a sample of radio sources at 1.4-GHz, combining 7 samples (3CRR, \citep{Laing83}, CoNFIG1-4 \citep{PaperI}, Combined EIS-NVSS Survey Of Radio Sources \citep[CENSORS;][]{Best03} and Lynx \& Hercules \citep{Rigby07}) covering a large range of flux densities (from $\SoneG \ge$0.5mJy for Lynx \& Hercules to $\SoneG \ge$3.5Jy for 3CRR). It includes FRI/FRII/Compact morphology classifications, optical identifications and redshift estimates. It contains 1114 sources and is 94.3\% complete for radio morphological classifications. Improvements to the catalogue are described in the following sections.

\subsection{GMRT Data}\label{GMRT}

In order to complete the spectral index coverage of the CoNFIG1-4 samples, GMRT data were obtained on July 4th, 2011 for 48 sources. They were observed over 9 hours in total (5 to 30 minutes on target depending on the source), with a central frequency of 591-MHz and a 33.3-MHz bandwidth divided into 256 channels. The source 3C 147 was used to calibrate flux densities and the data were reduced using standard \textsc{aips} procedures so as to reach a rms noise of $\sigma \approx$0.5 mJy/bm.

Flux densities were measured for all sources (Appendix~\ref{GMRTDat}), and the FRI/FRII morphology was confirmed for 4 sources with previously `possible' classification \citep{PaperII}: 4C 04.41 (C1-128, FRI), 4C 03.27 (C1-163, FRI), 4C 16.30 (C2-095, FRII) and 1438-0133 (C4-176, FRII). Contour plots for these sources are shown in Figure~\ref{NewMorph}.\\

\subsection{Spectral Index}\label{SpecInd}

Using the GMRT flux density measurements described in the previous section, in combination with the 1.4-GHz flux density data, previously unavailable spectral index values were computed for 46 sources. Spectral index determinations were also improved for a further 91 CoNFIG sources by including flux-density information from the VLA Low-frequency Sky Survey at 74MHz \citep[VLSS;][]{Cohen07} and from the Cosmic Lens All Sky Survey of radio sources at 8.4GHz \citep[CLASS;][]{Myers03}, or by recording values previously published. As specified by \cite{PaperII}, low-frequency spectral index determinations are preferred for our analysis, but high-frequency indices were used whenever the low-frequency ones were unavailable.

The revised and new spectral index values are presented in Appendix~\ref{AImprov}.\\

\subsection{K-band magnitude and redshifts}\label{ZandK}

Additional host-galaxy cross-identifications were performed using the UKIRT Infrared Deep Sky Survey DR9 \citep[UKIDSS;][]{Law07}. UKIDSS uses the UKIRT Wide Field Camera \cite[WFCAM;][]{Casali07} and a photometric system described in \cite{Hew06}. The pipeline processing and science archive are described in \cite{Hambly08}.

After visual inspection, K-band magnitudes (through a 2.0 arcsec aperture diameter) were retrieved for 190 CoNFIG sources (Appendix~\ref{UKIDSS}), including 20 new identifications (Appendix~\ref{UKIDSScoord}) and 48 extended radio sources with known spectroscopic redshifts. We computed a K-z relation appropriate to their magnitude determination (log(z) = 0.305K-5.319), which is in line with other K-z relations \citep[e.g.][]{Willott01,Brookes06}, and got the first redshift estimates for 25 FRI/II sources (including sources with optical identification but no previously available photometric or K-z redshift estimates). In addition, publications of new or updated catalogues \citep[e.g.][]{Rich09,Croom09} also allowed us to improve the redshift coverage of sources in the extended CoNFIG catalogue. The new redshift values are shown in Appendix~\ref{ZImprov}. Finally, redshift and spectral index information were updated for the CENSORS sample \citep{Brookes06} according to the work of \cite{Ker11}.\\

The improved catalogue includes a total of 760 extended sources (131 FRIs, 566 FRIIs and 63 uncertain) and 336 compact sources (not including 18 CSS sources), with 93.3\% spectral index completion (99.3\% for the four CoNFIG samples) and 82.9\% (spectroscopic or photometric) redshift coverage, making it one of the largest, most comprehensive databases of morphologically-classified radio sources and an important tool in the study of AGN space densities.

\subsection{The CoNFIG Local Sub-sample}

To investigate the nature of the physical processes behind the FR dichotomy, its relation to high/low excitation classification, and its dependence on environmental richness factor and host-galaxy luminosity, a sub-sample of local (z$\le$0.3) CoNFIG extended radio sources was compiled. The sub-sample contains 206 sources, comprising 73 FRIs, 103 FRIIs, 5 unclassified extended and 25 compact objects, and it is 99.5\% complete for spectral index and optical identification.\\


\section{The FR Dichotomy in the Local Universe}\label{FRdicho}  

\subsection{FRI/FRII LRLFs}  

\begin{table}
  \begin{center}
    \caption[]{\label{LRLFTab1}Local luminosity functions $\rho(P)$ data from Figure~\ref{LRLF} - FRI/FRII LRLF - and  Figure~\ref{HGLRLF} - HEG/LEG LRLF. P corresponds to the central 1.4 GHz luminosity of the bin (with $\Delta\logP$=0.4), and is given in $\uniP$. Space densities are in $\rm {Mpc^{-3}\Delta \logP^{-1}}$.} 
    \medskip
    \begin{tabular}{|c|cc|cc|}
      \hline
      {\bf P} & \multicolumn{4}{c}{\bf log$_{\bf 10}\bf (\rho)$}\\
                & \bf FRI        & \bf FRII       & \bf LEG        & {\bf HEG}       \\
      \hline
      22.2 & -4.66$\pm$0.23 &       -        & $-$5.21$\pm$0.30 &         -        \\
      22.6 &   -            & -5.66$\pm$0.23 &         -        & $-$5.67$\pm$0.30 \\
      23.0 & -5.06$\pm$0.14 & -6.07$\pm$0.14 & $-$5.19$\pm$0.16 &         -        \\
      23.4 & -5.29$\pm$0.10 & -5.86$\pm$0.10 & $-$5.25$\pm$0.24 &         -        \\
      23.8 & -5.69$\pm$0.09 & -6.05$\pm$0.09 & $-$5.53$\pm$0.17 & $-$6.33$\pm$0.52 \\
      24.2 & -6.23$\pm$0.10 & -6.19$\pm$0.10 & $-$6.04$\pm$0.23 & $-$6.64$\pm$0.59 \\
      24.6 & -6.79$\pm$0.11 & -6.46$\pm$0.11 & $-$6.68$\pm$0.43 & $-$6.51$\pm$0.31 \\
      25.0 & -7.89$\pm$0.23 & -6.91$\pm$0.23 & $-$7.47$\pm$0.63 & $-$7.06$\pm$0.32 \\
      25.4 & -8.28$\pm$0.23 & -7.36$\pm$0.23 & $-$7.92$\pm$0.65 & $-$7.41$\pm$0.29 \\
      25.8 & -8.83$\pm$0.30 & -8.35$\pm$0.30 & $-$8.36$\pm$0.20 & $-$7.99$\pm$0.14 \\
      26.2 &  -             & -8.83$\pm$0.30 &      -           &       -          \\
      \hline
    \end{tabular}
  \end{center}
\end{table}

Using the updated extended CoNFIG catalogue, the local radio luminosity functions (LRLF) were computed using the $\rm 1/V_{max}$ technique for z$\le$0.3 (with $\logP \ge 22.0 \uniP$), in which, for each P-z bin, the space density is given by:
\begin{eqnarray}\label{EqVmax}
  \rho = \sum_{i=1}^N\frac{1}{V_i}\ \ \ \ \ \ \ \ \ \ \ 
  \sigma^2 = \sum_{i=1}^N\frac{1}{V_i^2}
\end{eqnarray}
where $\rm V_i$ is the largest volume in which the source could be observed in bin {\it i}.\\

Comparing the FR LRLFs presented here (Figure~\ref{LRLF} \& Table \ref{LRLFTab1}) with Figure~12 of \cite{PaperII}, the improvement in CoNFIG allowed for a better definition of the LRLFs. In particular, for FRIs, the space density determinations extend to higher luminosities, while at lower luminosities, the FRII LRLF seems to plateau for $\logP \le \rm 23.8 \ \uniP$.\\

\subsection{High/low excitation galaxies}\label{Excitation}

\begin{figure*}
  \begin{minipage}{16.0cm}
    \begin{minipage}{7.5cm}
      \centerline{
        \includegraphics[angle=270,scale=0.35]{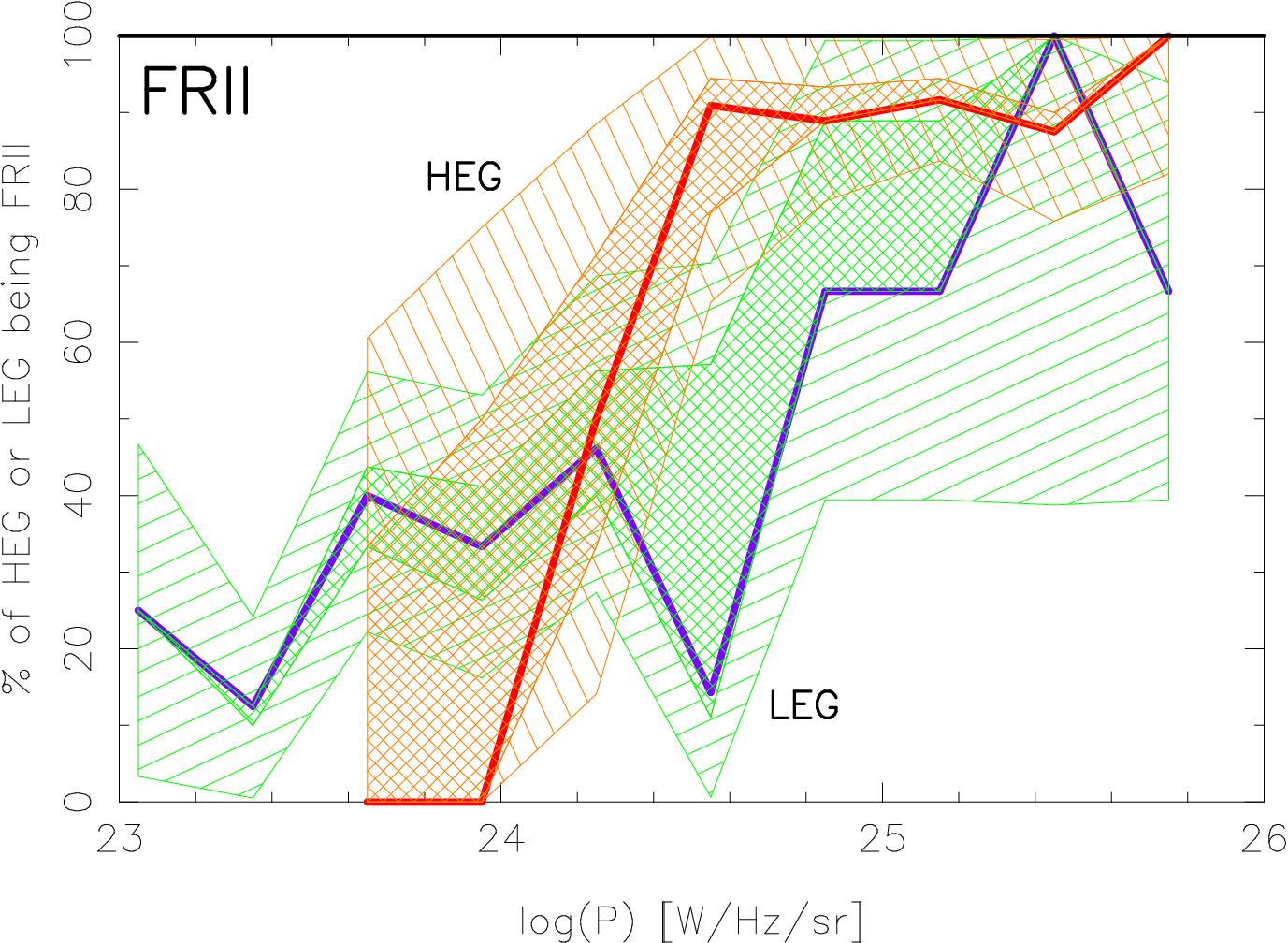}}
    \end{minipage}
    \hfill
    \begin{minipage}{7.5cm}
      \centerline{
        \includegraphics[angle=270,scale=0.35]{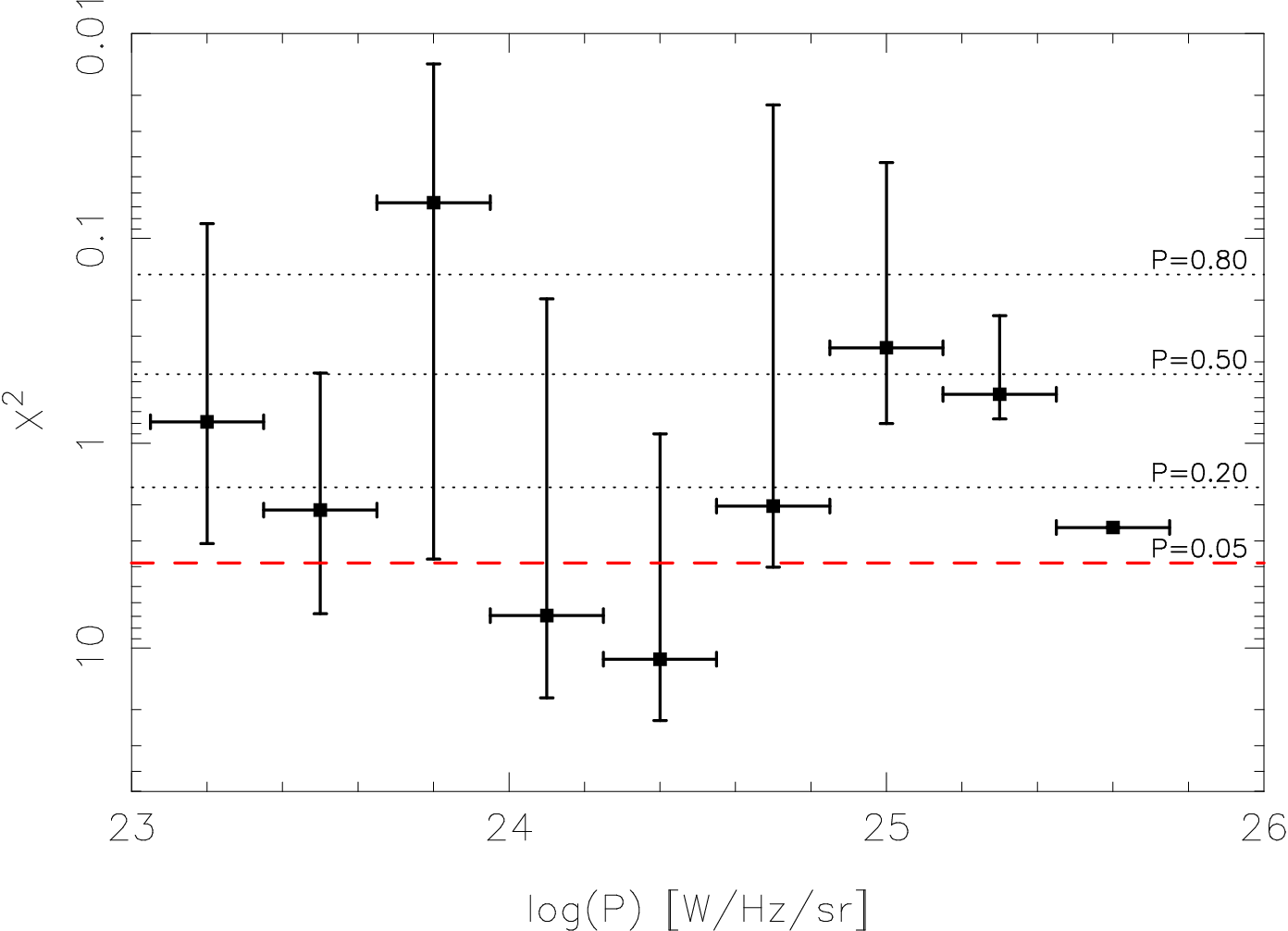}}
    \end{minipage}
  \end{minipage}
  \caption[]{\label{FRRatio} Left: Percentage of HEGs (thick red line) and LEGs (thick purple line) being FRII in the CoNFIG local sub-sample (excluding sources with no HEG/LEG classification). The luminosity bins are $\Delta$logP=0.3 wide. The cross-hatched regions correspond to the minimum and maximum possible values of the ratios when including sources with unidentified spectral type, and the hatched regions include errors in these limits depending on the number of sources in each bins following Poisson statistics.\\
Right: Result of the Pearson chi-square test performed on the FRI/FRII HEG/LEG samples for each luminosity bin. For comparison, $\chi^2$ values corresponding to a probability P=0.2, 0.5 and 0.8 that radio morphology is independent of excitation are displayed in dotted lines. The dashed line represent the value of $\chi^2$ for which P=0.05, the lowest acceptable probability for which the distributions are independent accepted here. For each luminosity bin, the range of possible $\chi^2$ values when including unclassified sources is determined based on the minimum and maximum values of $\chi^2$ in each of the following extreme cases: (i) no unclassified sources are taken into account; (ii) all unclassified sources are LEG; (iii) all unclassified sources are HEG; (iv) all FRI unclassified sources are LEG while all FRII unclassified sources are HEG; (v) all FRI unclassified sources are HEG while all FRII unclassified sources are LEG. These error bars indicate the degree of uncertainty imposed by the lack of complete classification.} 
\end{figure*}

In this work, HEG/LEG classification was determined by measuring the [O{\small III}] ($\lambda_{\rm [O{\small III}]}$ = 5007\AA) and [O{\small II}] ($\lambda_{\rm [O{\small II}]}$ = 3727\AA) lines, and following the definitions of \cite{Jack97}: sources with rest-frame [O{\small III}] equivalent width $<$ 1nm and/or [O{\small II}]/[O{\small III}] $>$ 1 were classified as LEG, other sources being classified as HEG. If no [O{\small III}] line was detected in the spectrum (in which a EW$\sim$1nm line would be otherwise detected), the source was considered to be low excitation.

For the CoNFIG local sub-sample, we found 88 LEGs (including 49 FRIs and 29 FRIIs) and 70 HEGs (including 11 FRIs and 47 FRIIs). The 48 other sources (including 13 FRIs and 27 FRIIs) did not have spectra available to determine the excitation level of the host galaxy. The HEG/LEG classification is shown in Appendix~\ref{Local}.\\

Looking at the host galaxies properties of sources with no HEG/LEG classification available, no major systematic offsets in magnitude or other properties were observed compared to other radio sources of the same redshifts and radio fluxes. The local radio luminosity function was computed for both HEGs and LEGs following Equ.~\ref{EqVmax}, with these unclassified sources were therefore considered to be a random sub-sample and were taken into account by correcting each luminosity bins of the LRLFs by a factor:
\begin{eqnarray}\label{Corrfact}
  F=1+\frac{\left.\sum_{i=1}^N\frac{1}{V_i} \right|_{unclass.}}{ \left.\sum_{i=1}^N\frac{1}{V_i} \right|_{classified}}
\end{eqnarray}
The resulting LRLFs are shown in Figure~\ref{HGLRLF}. We see that for both HEGs and LEGs, the data cover the full range of radio luminosities studied ($\rm 22.0 \le \logP \le$ 26.0 $\uniP$), and they agree well with the work of \cite{Best12}, indicating that the inclusion of sources with no HEG/LEG classification was properly done. Indeed, in regions of the LRLF where the space density of HEGs and LEGs differ by an order of magnitude, if too many unclassified sources had been added to the less-dominant population, they could have produced a factor few increase on that LRLF. We do find a higher space density of HEGs in our sample for $\logP \ge$ 24 $\uniP$ relative to \cite{Best12}, but no deficiency in LEGs. Part of this is caused by Best \& Heckman's exclusion of quasars which, although a small proportion of the overall sample, are a significant fraction of high power HEGs. Nevertheless, a small excess is still present when applying the \cite{Best12} selection criteria, suggesting that optically selected samples, such as SDSS, might be under-sampling high-power HEGs.

\subsection{Cluster richness}\label{Env}  

Cluster richness for each source was determined using the method of \cite{Wing11}, in which the richness factor N$^{-19}_1$ corresponds to the corrected number of SDSS galaxies with absolute magnitudes brighter than M$_{r}$ = $-$19 within a 1.0\,Mpc radius of the radio source. The corrected galaxy count is obtained by measuring the total number of sources in the 1.0\,Mpc-radius disk and subtracting a background count, measured from a shell of inner and outer radii 2.7 and 3.0 Mpc respectively.\\

When SDSS data were unavailable (20.4\% of the local sample), SuperCosmos Sky survey R-band (28 sources) and EIS Patch-D I-band (14 CENSORS sources) data were used. The {\it r}-band to R-band and {\it r}-band to I-band magnitude limit conversion were determined from sources in the CoNFIG local sample with both data available, and are given as: 
\begin{eqnarray}
  \rm R = {\it r} - 0.64 & (\sigma_{rms}=0.09) \label{EqRRr}\\
  \rm I = ({\it r} - 0.46) - 0.75 & (\sigma_{rms}=0.3) \label{EqRIr}
\end{eqnarray}
with I = {\it i} - 0.75 as the standard conversion from \cite{Wind91}.\\ 

According to \cite{Wing11}, a cluster-richness of N$^{-19}_1 \le 20$ likely corresponds to a poor environment, while N$^{-19}_1 \ge 40$ corresponds to a rich cluster. It was thus decided to use N$^{-19}_1 = 30$ to differentiate between poor and rich environments.
Values of N$^{-19}_1$ for sources in the local CoNFIG sub-sample are shown in Appendix~\ref{Local}.

\section{Results \& Discussion}\label{Results}  

\begin{table}
  \begin{center}
    \caption[]{\label{EnvDatTab}Environmental parameters for each of the populations (FRI,FRII, HEG, LEG and combinations) considered in this work.} 
    \medskip
    \begin{tabular}{|l|rrr|rc|}
      \hline
      Type & \multicolumn{3}{c}{Number of sources} & \multicolumn{2}{c}{Richness} \\
      &\multicolumn{1}{c}{tot.} &\multicolumn{1}{c}{poor} &\multicolumn{1}{c}{rich} & \multicolumn{1}{c}{mean} & \multicolumn{1}{c}{median} \\
      &&&& \multicolumn{1}{c}{($\mu\pm\Delta\mu$)} & \multicolumn{1}{c}{M} \\
      \hline
      FRI      &  73 & 36 & 37 & 31.9$\pm$ 7.7 & 29.8 \\
      FRII     & 103 & 77 & 26 & 20.3$\pm$ 3.4 & 14.9 \\
      \hline
      HEG      &  58 & 56 & 14 & 19.8$\pm$ 5.3 & 15.1 \\
      LEG      &  78 & 48 & 40 & 31.9$\pm$ 7.1 & 29.8 \\
      Unk.     &  40 & 35 & 13 & 19.6$\pm$ 4.4 & 15.7 \\
      \hline
      FRI-HEG  &  11 &  9 &  2 & 14.3$\pm$10.3 & \phantom{1}3.6 \\
      FRI-LEG  &  49 & 21 & 28 & 36.3$\pm$11.0 & 35.8 \\
      FRII-HEG &  47 & 38 &  9 & 21.0$\pm$ 6.1 & 15.4 \\
      FRII-LEG &  29 & 18 & 11 & 24.6$\pm$ 4.5 & 14.3 \\
      \hline
    \end{tabular}
  \end{center}
\end{table}
\begin{figure}
  \centerline{
    \includegraphics[angle=270,scale=0.35]{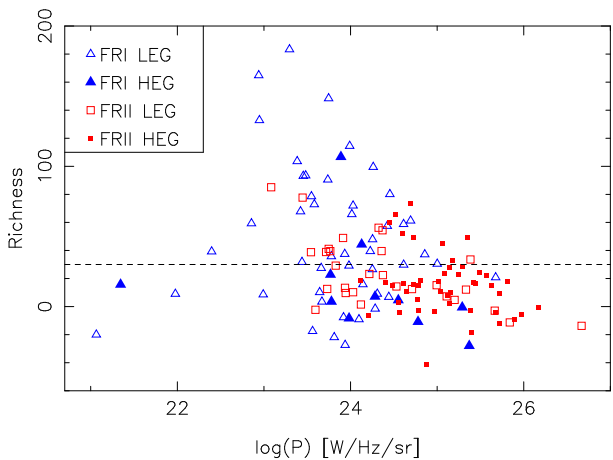}}
  \caption[]{\label{RichLum}Richness factor as a function of radio power for FRI (blue triangles) and FRII (red squares) HEGs (filled symbols) and LEGs (open symbols). The dashed line correspond to N$^{-19}_1 = 30$, the limit between poor and rich cluster as defined in \S\ref{Env}.}
\end{figure}

\subsection{HEG/LEG}\label{HLRes}

The possibility that FR types depend on the distinct accretion mode inside the central SMBH is examined by looking at the probability of a HEG/LEG being of a given FR type.\\

The fractions of HEG and LEG being FRII, as a function of radio power, are displayed in the left panel of Figure~\ref{FRRatio}. The two distribution overlap within the errors, which include both uncertainty due to sources with no HEG/LEG classification (17.6\% of FRI and 26.5\% of FRIIs) and Poisson statistics dependent on the number of sources in each luminosity bin considered. It appears that Poisson errors are the main source of uncertainty here.

A Pearson chi-square test, including Yate's correction when appropriate, was performed on the FRI/FRII HEG/LEG samples for each luminosity bin (right panel of Figure~\ref{FRRatio}). The degree of uncertainty imposed by the lack of complete classification is indicated here by including sources with no excitation classification in different categories and is represented as error bars. In most luminosity bins (apart from $23.95 \le \logP \le 24.55 \uniP$), the probability of radio morphology being independent of excitation is greater than 5\%, and up to P$_{FR-H/L}>$80\% in a third of cases. For the intermediate luminosity range singled out above, there is some indication that there might be a difference at the 5\% confidence limit. However, the idea that there's a dependence on excitation state at intermediate luminosities that isn't present at other luminosities seems unphysical, in particular when considering the relatively low confidence level of the difference. Especially when considering the potential influence of sources without excitation classification, it appears possible that FRI/FRII are independent of HEG/LEG type over the whole range of luminosity considered. \\

Thus, based on the above results, the null hypothesis that, at given radio luminosity, FR morphology is independent of the accretion-mode of the black hole, can not be ruled out.

\subsection{Environmental influences}

\begin{figure*}
  \begin{minipage}{16.0cm}
    \begin{minipage}{7.5cm}
      \centerline{
        \includegraphics[angle=270,scale=0.35]{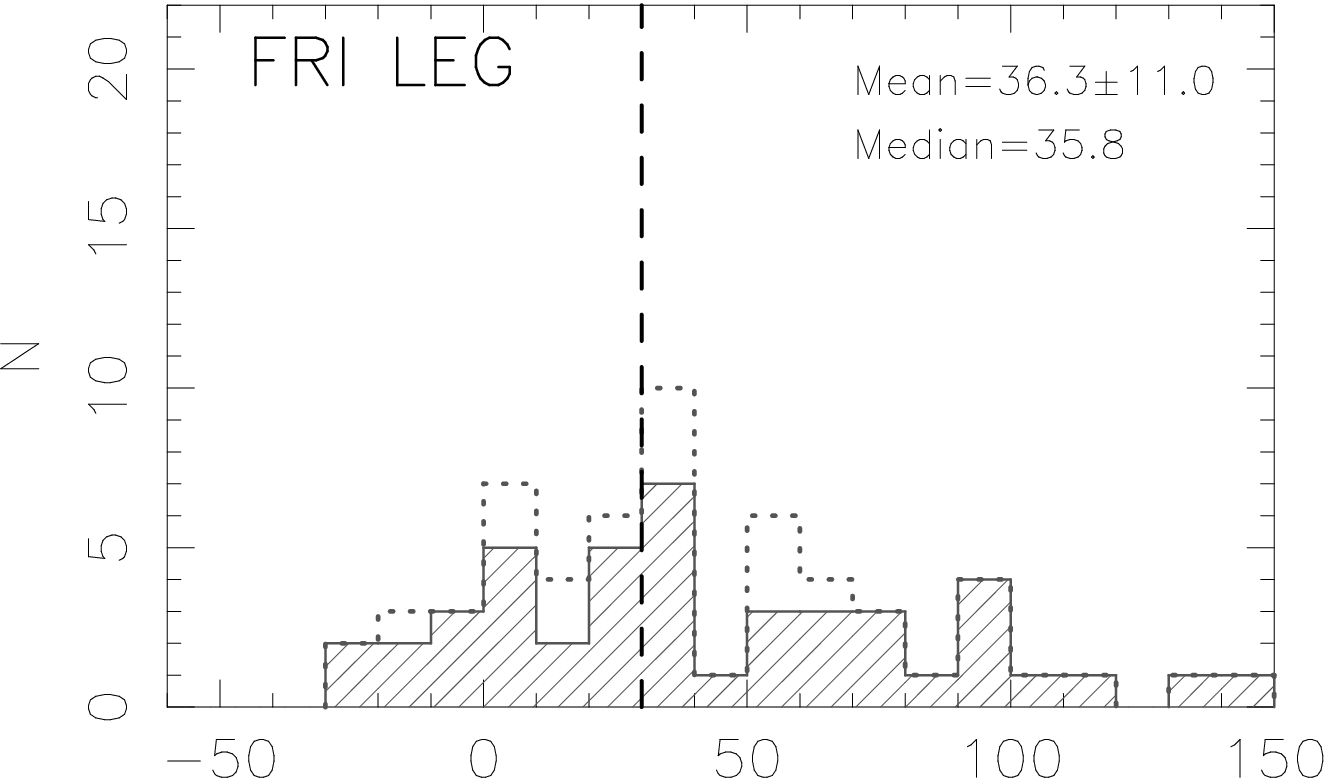}}
    \end{minipage}
    \hfill
    \begin{minipage}{7.5cm}
      \centerline{
        \includegraphics[angle=270,scale=0.35]{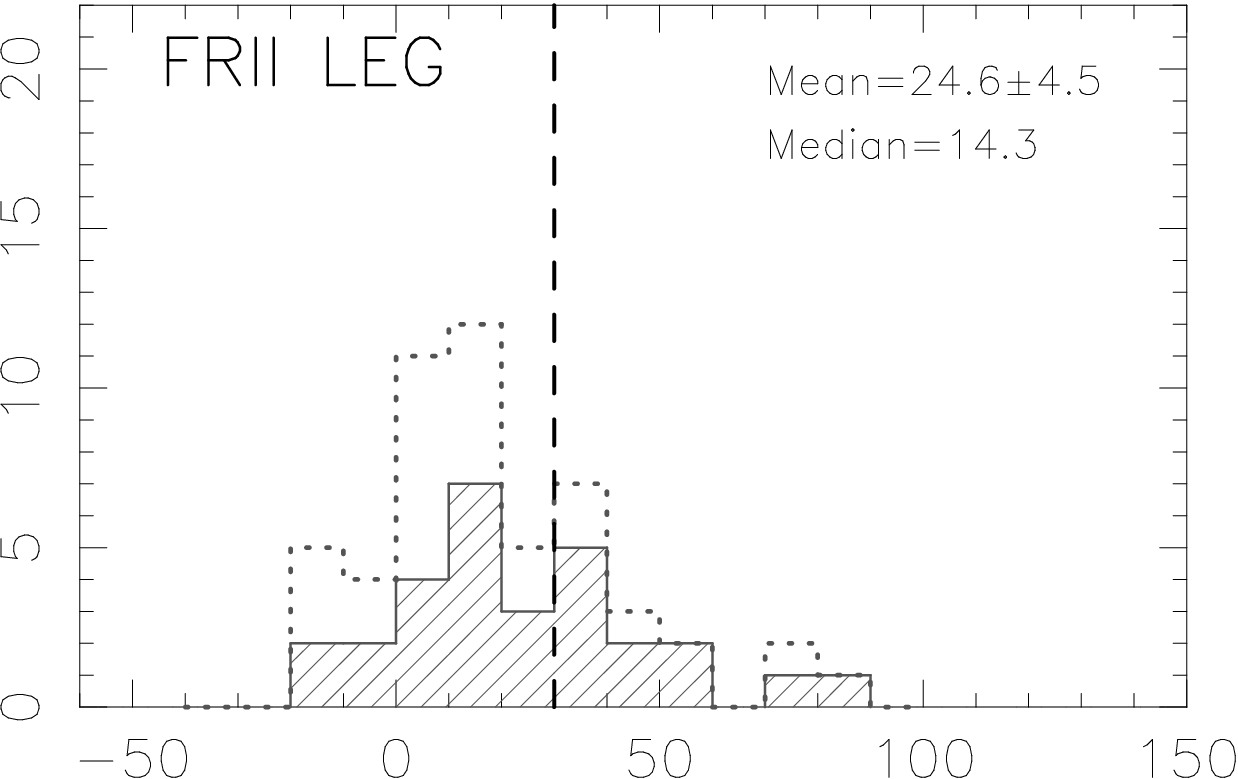}}
    \end{minipage}
  \end{minipage}
  \vfill
  \vspace{2mm}
  \begin{minipage}{16.0cm}
    \begin{minipage}{7.5cm}
      \centerline{
        \includegraphics[angle=270,scale=0.35]{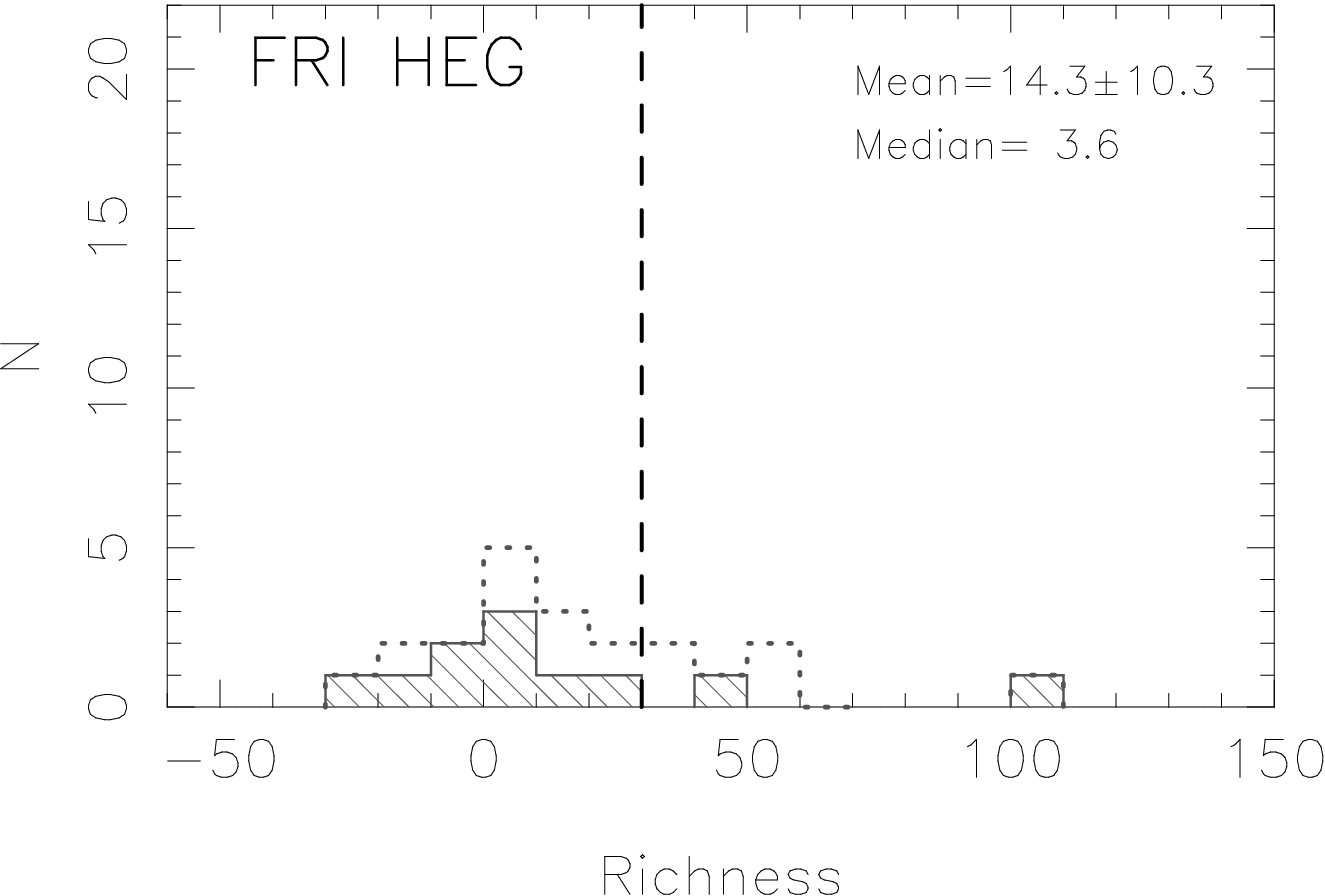}}
    \end{minipage}
    \hfill
    \begin{minipage}{7.5cm}
      \centerline{
        \includegraphics[angle=270,scale=0.35]{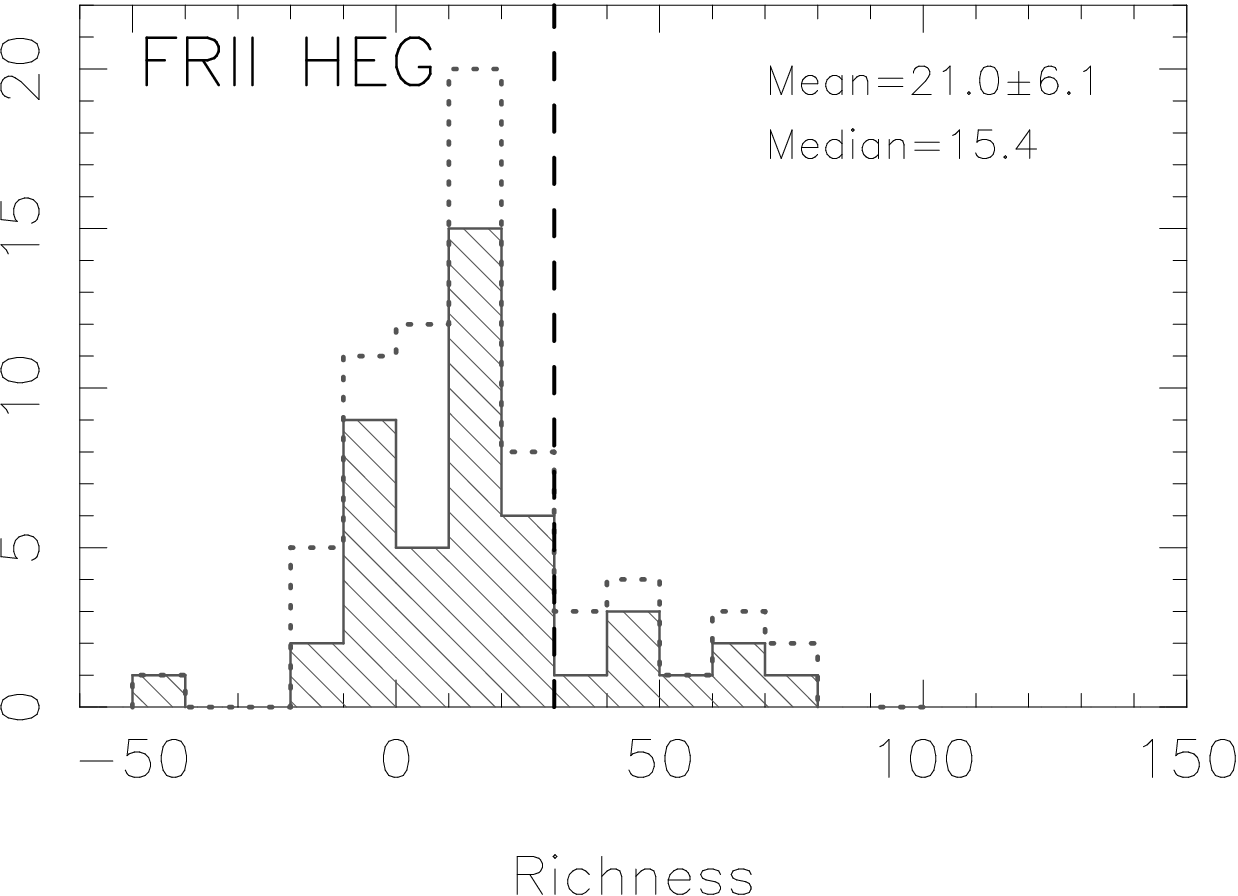}}
    \end{minipage}
  \end{minipage}
  \caption[]{\label{FR_HL_PR}Richness distribution for FRI (left) and FRII (right) LEG (top) and HEG (bottom) sources in the CoNFIG local sub-sample. The dashed line correspond to N$^{-19}_1 = 30$, the limit between poor and rich cluster as defined in \S\ref{Env}. The richness distribution taking into account sources for which HEG/LEG classification was not possible are represented as dotted histograms. Mean, error on mean and median richness (without unclassified sources) are quoted for each distribution.}
\end{figure*}
\begin{figure*}
  \begin{minipage}{16.0cm}
    \begin{minipage}{7.5cm}
      \centerline{
        \includegraphics[angle=270,scale=0.35]{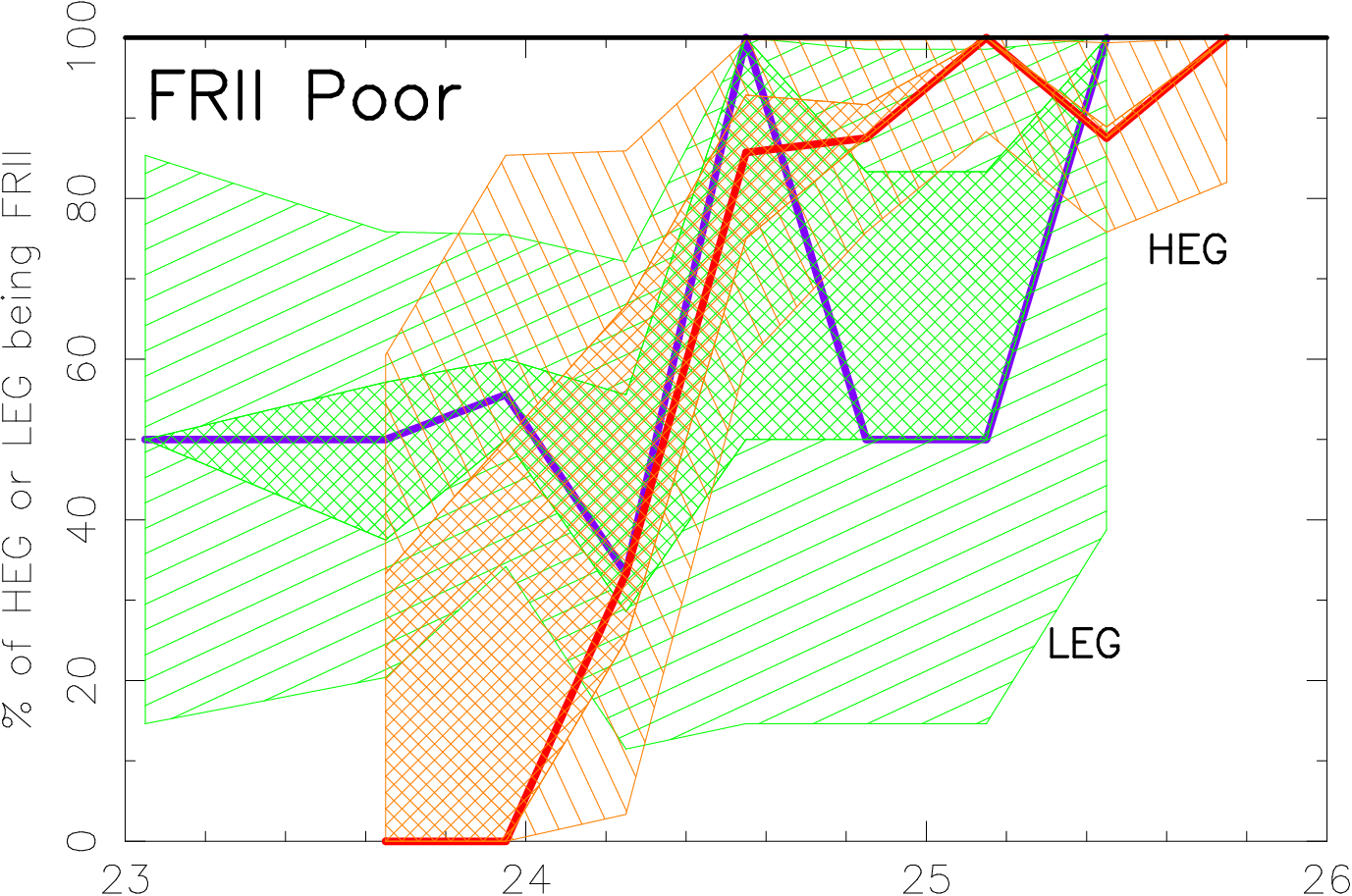}}
    \end{minipage}
    \hfill
    \begin{minipage}{7.5cm}
      \centerline{
        \includegraphics[angle=270,scale=0.35]{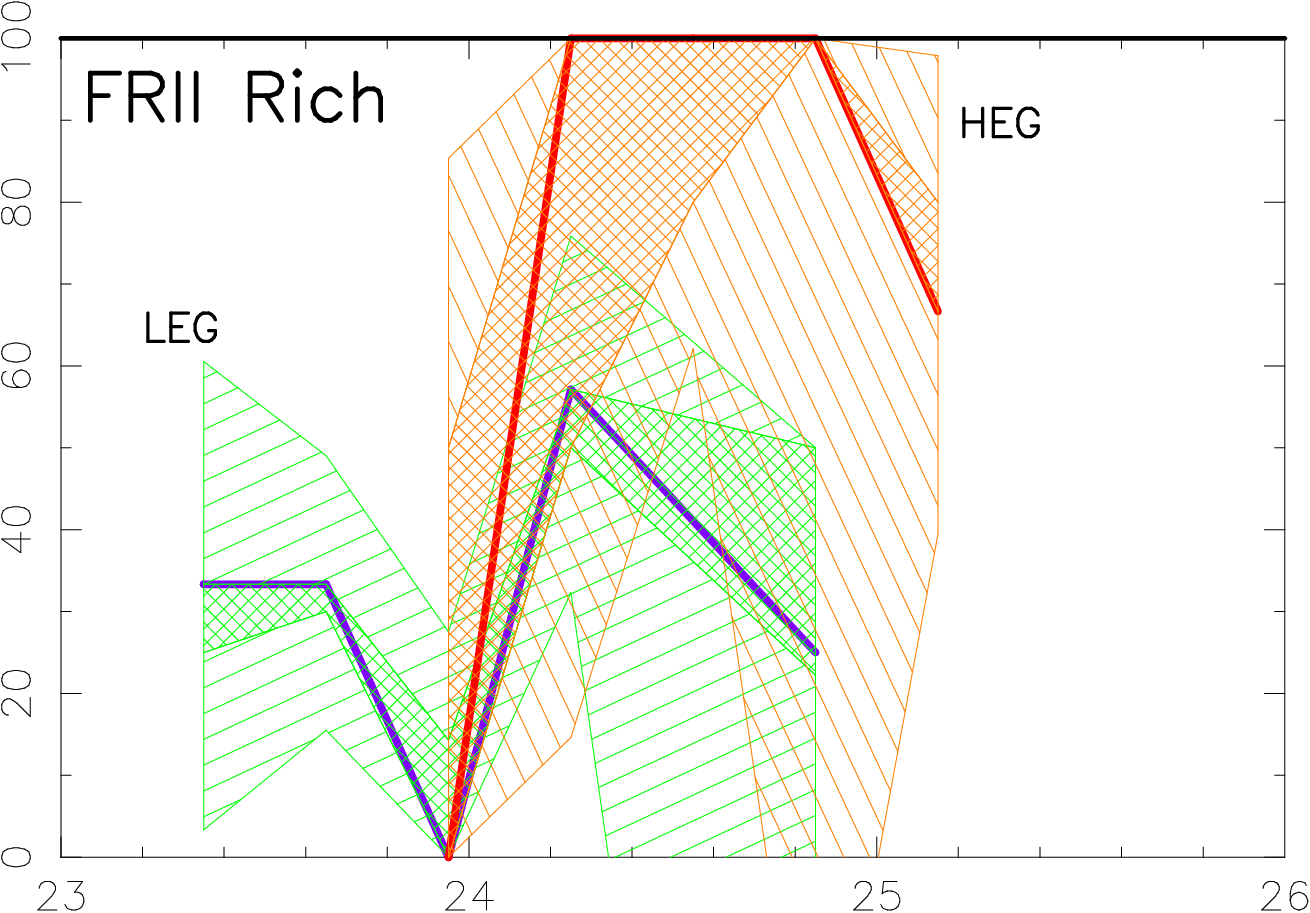}}
    \end{minipage}
  \end{minipage}
  \vfill
  \vspace{2mm}
  \begin{minipage}{16.0cm}
    \begin{minipage}{7.5cm}
      \centerline{
        \includegraphics[angle=270,scale=0.35]{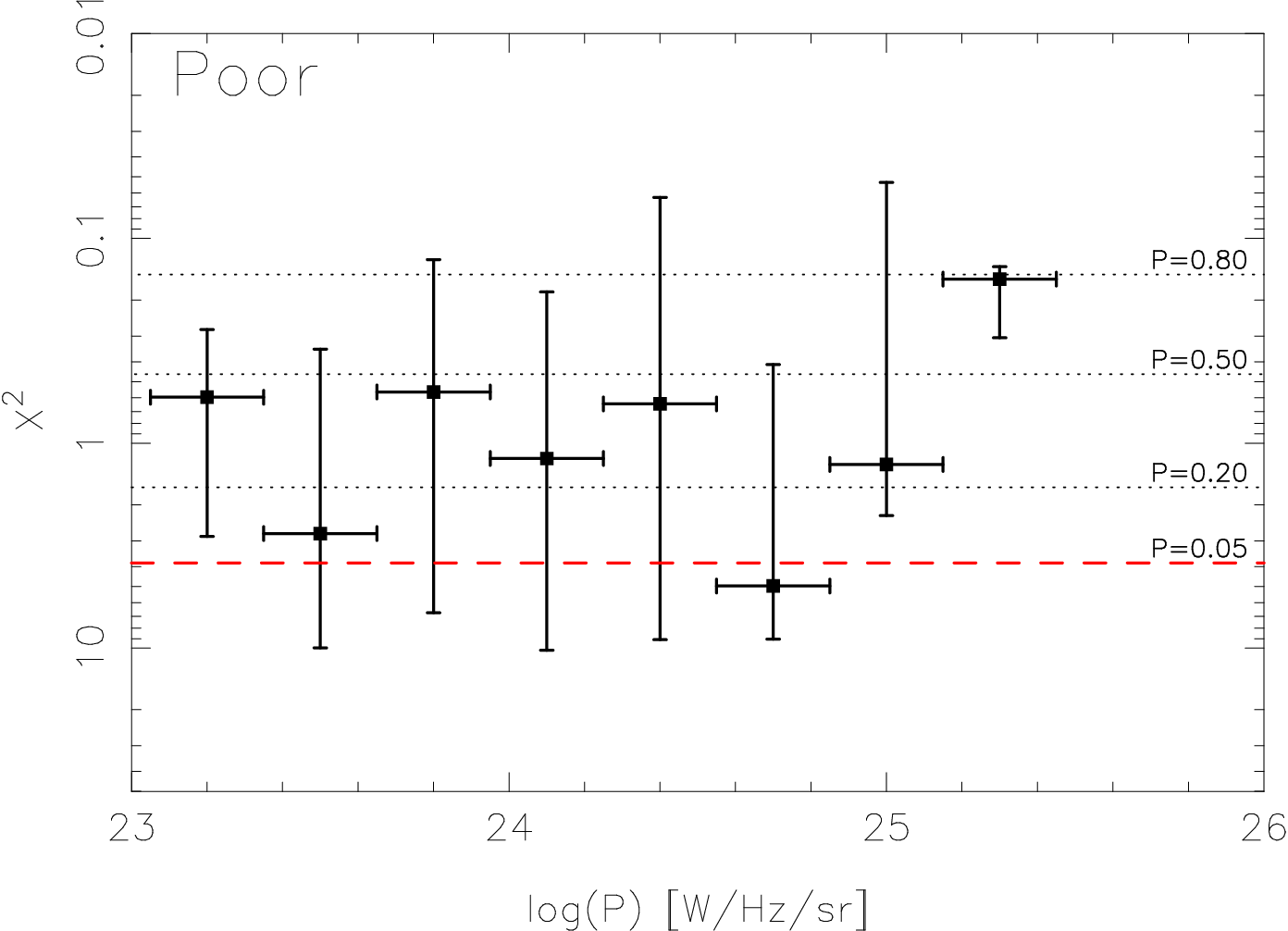}}
    \end{minipage}
    \hfill
    \begin{minipage}{7.5cm}
      \centerline{
        \includegraphics[angle=270,scale=0.35]{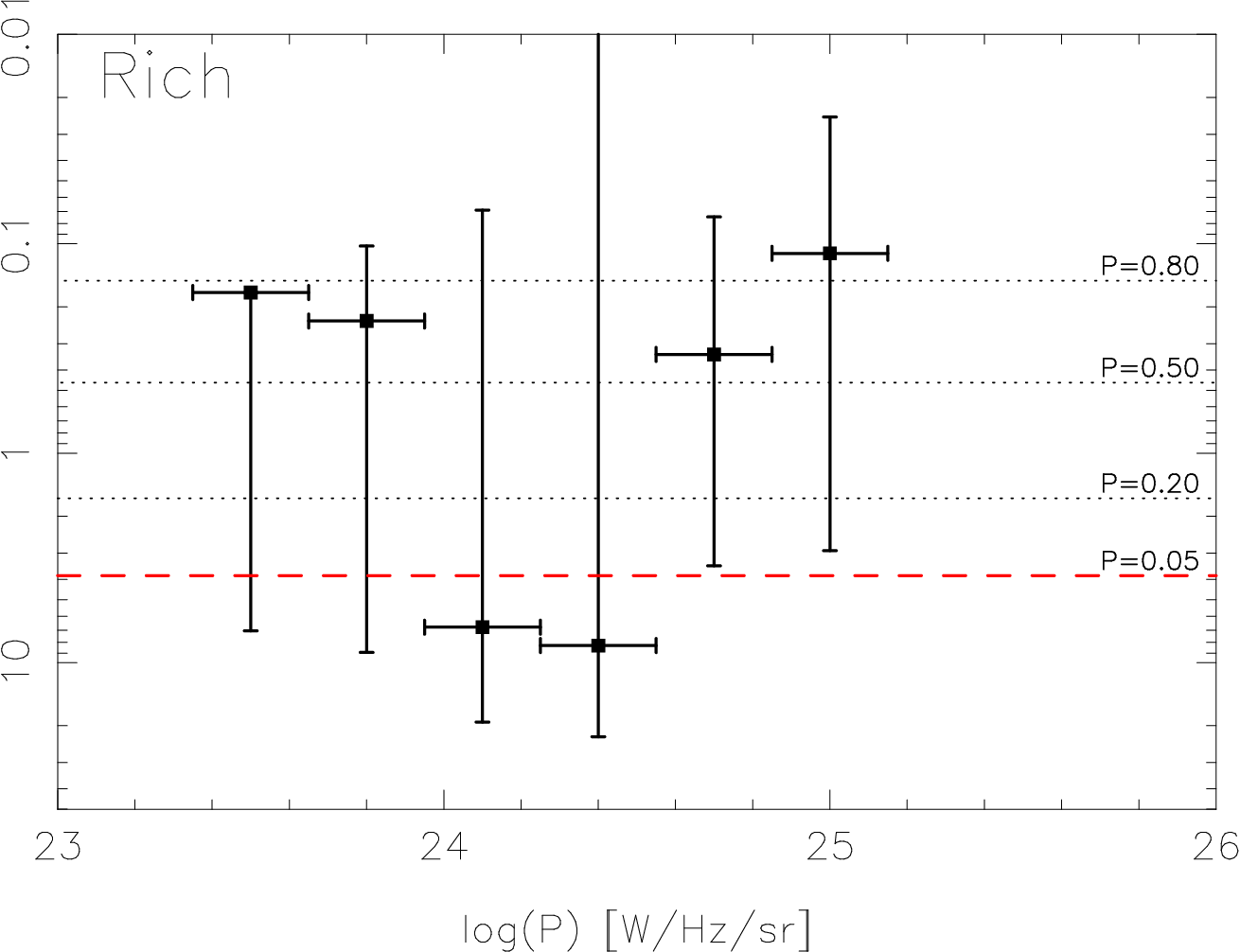}}
    \end{minipage}
  \end{minipage}
  \caption[]{\label{FRRatio_env} Top: Percentage of HEGs and LEGs being FRII in a poor (left) or rich (right) cluster in the CoNFIG local sub-sample (excluding source with no HEG/LEG classification), with identical references as Figure~\ref{FRRatio}. The last point of the rich LEG ratios was computed in a bin of size $\Delta$logP=0.9 to increase the number of sources in the bin.\\
Bottom: Result of the Pearson chi-square test performed on the FRI/FRII HEG/LEG samples for each luminosity bin in a poor (left) or rich (right) cluster, with identical references as figure~\ref{FRRatio}.} 
\end{figure*}

\subsubsection{Cluster richness}\label{ClusterRes}

The idea that, no matter how they are produced, jets will behave differently depending on the cluster environment they encounter is a possible explanation for the different FR morphologies, independently of excitation types. Environmental statistics for each of the populations considered in this work are presented in Table~\ref{EnvDatTab}, while Figure~\ref{RichLum} shows the distribution of richness factor with respect to radio powers. Note that, based on Figure~\ref{RichLum}, it seems that radio power offsets between FRI and FRII sources will not be a strong source of bias in the following analysis.\\

Looking at the environmental difference between FRI and FRII only, a Pearson chi-square test leads to $\chi^2_{\nu=1}=9.07$, rejecting the hypothesis that FR morphology and environment parameters are independent with probability P$_{FR}$=0.990. With median richness $M_I$=29.8 and $M_{II}$=14.9, it appears that FRI sources tend to be located in richer clusters than FRIIs, as previously stated by \cite{Zirb97} and \cite{Prest88}.

Focusing on environmental differences between HEGs and LEGs, it can be seen in Figure~\ref{RichLum} that HEGs are found almost exclusively in low-density environments, with median richness $M_H$=15.1. In contrast, LEGs are found in a wider range of densities. A Pearson chi-square test is performed on samples of HEGs and LEGs in poor and rich clusters (ignoring unclassified sources) leading to $\chi^2_{\nu=1}=14.23$, rejecting the hypothesis that excitation mode and environment parameters are independent with probability P$_{H/L-Rich}$=0.998. The dependence of the accretion mode on the environment can possibly be explained by the feeding mechanism associated with each type. Indeed, these results are consistent with HEGs being the result of interactions or mergers (which tend to occur in groups, with lower densities than clusters), while the gas supply of LEGs originates from the cooling out of either the host galaxy itself (possible in both rich and poor environments) or the cluster halo (requiring high densities).\\

Since both FRI/FRII and HEG/LEG sources show significant environmental influence, and since there are large overlaps in FRI-HEG and FRII-LEG populations, it is essential to test whether both of these relations are independently valid, or whether one is being driven by the other. The richness distribution of FRI/FRII HEGs/LEGs is shown in Figure~\ref{FR_HL_PR}.

To further look into a possible FR morphology - excitation mode dependence (or lack thereof), a Kolmogorov-Smirnov test performed for four comparative cases: FRI HEGs and LEGs, FRII HEGs and LEGs, LEG FRIs and FRIIs and HEG FRIs and FRIIs. The probabilities that the considered samples are drawn from the same distribution are P$_{\rm I-H/L}$=0.02, P$_{\rm II-H/L}$=0.30, P$_{\rm L-I/II}$=0.02 and P$_{\rm H-I/II}$=0.27 respectively. The probability P$_{\rm I-H/L}$ seems to indicate that, for FRI sources, HEGs and LEGs show a difference in richness. There is hence an environmental dependence on HEG/LEG not driven by FR morphology. Similarly, the low value of P$_{\rm L-I/II}$ shows that there exists an environmental dependence on FRIs/FRIIs not driven only by the accretion mode of the source. Overall, FRI-LEGs stand out as the only class with a substantial number of sources located in high density environments. When restricting the test to a narrow luminosity range (23.5$\le \logP \le$25.0$\uniP$), thus reducing as much as possible the effects of any trends with luminosity, similar results (P$_{\rm I-H/L}$=0.01, P$_{\rm II-H/L}$=0.97, P$_{\rm L-I/II}$=0.02 and P$_{\rm H-I/II}$=0.32) were found, verifying that no biases are caused by underlying correlation between luminosity and environment.

Finally, a similar analysis to the one presented in \S\ref{HLRes} is performed, looking at the fractions of HEG and LEG being FRII in poor (N$^{-19}_1 \le 30$) and rich (N$^{-19}_1 > 30$) environments (top panels of Figure~\ref{FRRatio_env}). A Pearson chi-square test, including Yate's correction when appropriate, was performed in each case (bottom panels of Figure~\ref{FRRatio_env}). For poor clusters, the probability of radio morphology being independent of excitation is greater than 5\% for most luminosity bins, even when including sources with no excitation classification. The results are similar for sources in rich clusters. Overall, this suggests that radio morphology is not fully determined by the combination of accretion mode and cluster density.

\subsubsection{Host galaxy}

\begin{figure*}
  \begin{minipage}{16.0cm}
    \begin{minipage}{7.5cm}
      \centerline{
        \includegraphics[angle=270,scale=0.31]{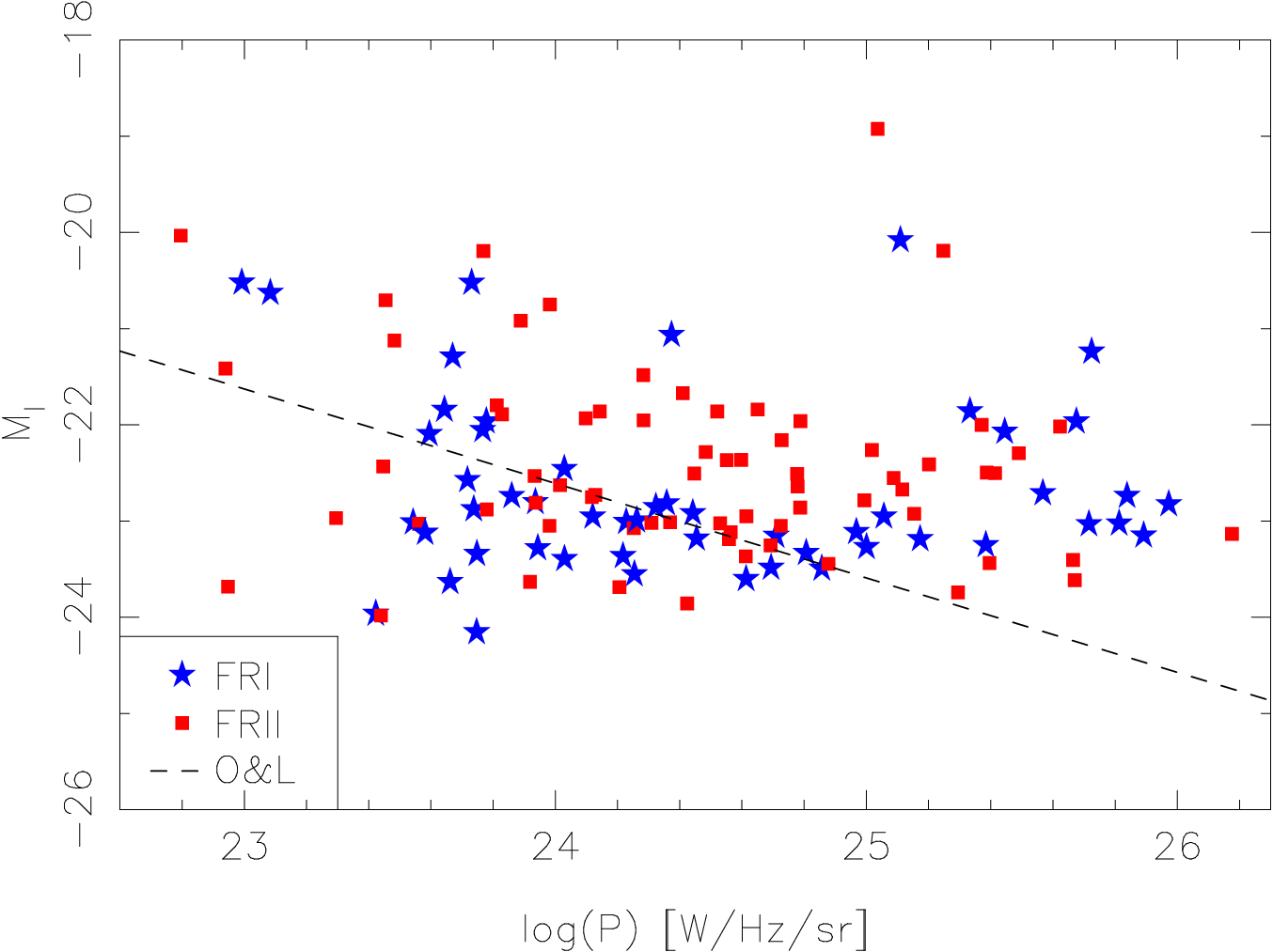}}
    \end{minipage}
    \hfill
    \begin{minipage}{7.5cm}
      \centerline{
        \includegraphics[angle=270,scale=0.31]{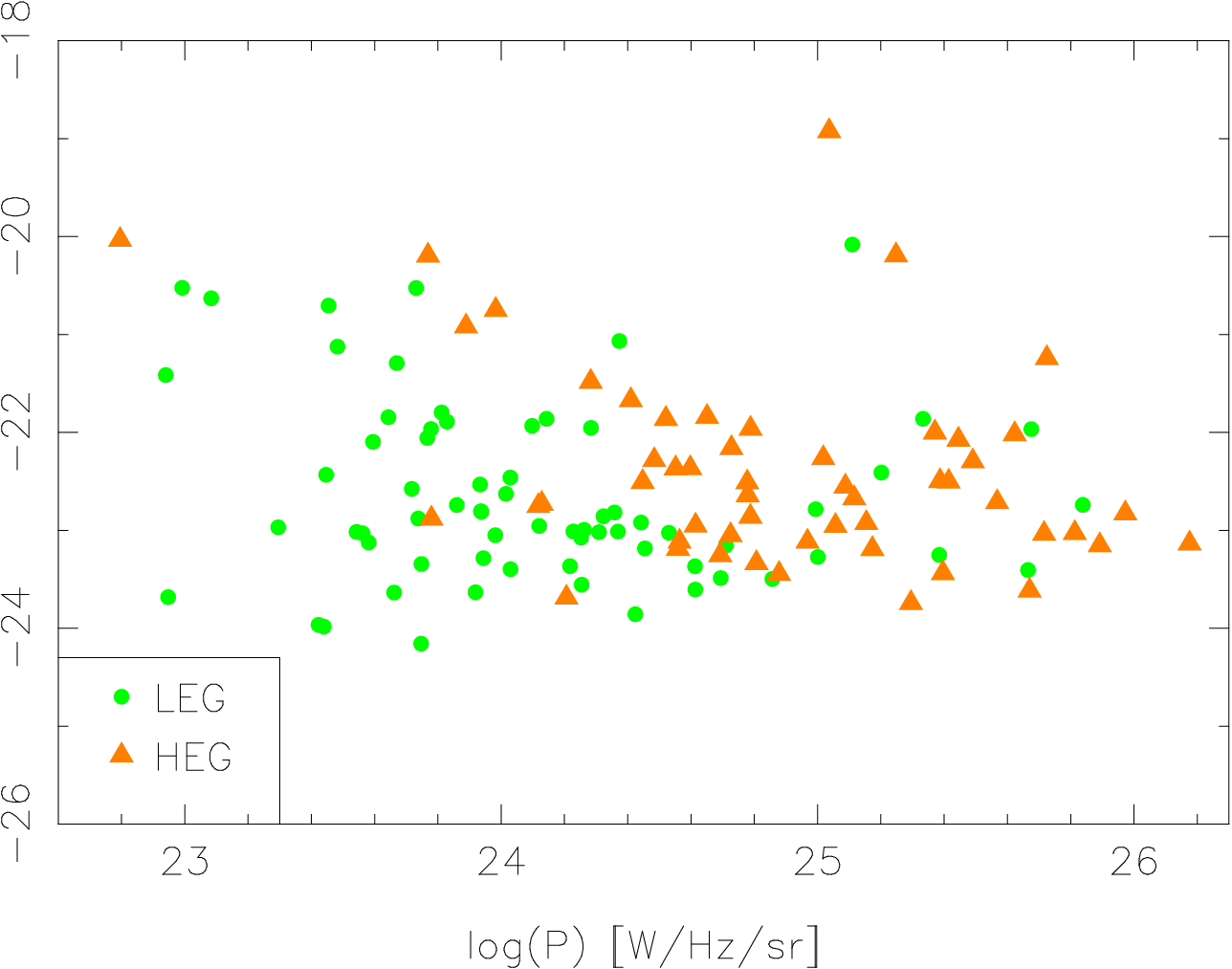}}
    \end{minipage}
  \end{minipage}
  \caption[]{\label{LumLum}Optical I-band vs. radio luminosity for FRI and FRII (left panel) and LEGs and  HEGs (right panel), excluding quasar sources. For the FRI/II plot, the \cite{Led96} relation is also displayed.}
\end{figure*}

According to the previous results, the disruption of the jets leading to the different FR types, although having some dependence on large cluster scale, still show a clear overlap of environment densities. Another factor considered in this work is that the disruption occurs on the scale of the host galaxy. This was suggested by \cite{Led96}, who found that the FR division is a function of both optical and radio luminosity while considering sources in the 3CR sample. However, several recent studies \citep{Best09,Wing11}, based on other independent samples, failed to replicate the sharp division found between FR populations, finding a large overlap around the Ledlow \& Owen divide. Looking at the M$\rm _I$-$\logP$ plots for the CoNFIG local sub-sample presented in Figure~\ref{LumLum}, it is apparent that the result of \cite{Led96} does not hold for the CoNFIG local sub-sample, even when considering the different intrinsic and extrinsic parameters. This implies that radio galaxies of different FR type are not hosted by significantly different galaxies. In contrast, HEGs and LEGs appear to occupy different regions of the M$\rm _I$-$\logP$ space. However, this separation is mostly radio-power driven. Note that in the luminosity range 24.0$\le \logP \le$25.0$\uniP$, the transition luminosity range between the HEG/LEG and FRI/FRII luminosity functions, some trends can be observed, such as FRII being hosted by galaxies extending to lower optical luminosity than FRIs at a given radio power. The significance of these trends is weak, and they are not present at other radio luminosities, but it is notable that this is the same radio luminosity range in which a potential difference was also observed in Figure~\ref{FRRatio}. A larger sample will be required to establish whether these differences are real.

\section{Discussion and conclusion}

In this paper, a sub-sample of local (z$\le$0.3) sources from the extended CoNFIG catalogue was used to determine the nature of the relation between morphology and accretion mode in radio galaxies, including environmental parameters. High/low-excitation status were determined for each source by retrieving spectroscopic data, in the majority from SDSS, and comparing the characteristics of [O{\small II}] and [O{\small III}] lines. Cluster richness factors were computed for each source based on the method presented in \cite{Wing11}, from SDSS, SSS or EIS photometric data. The local sub-sample contains 206 sources, including 74 FRIs and 102 FRIIs and is 100\% and 76.7\% complete for cluster density and HEG/LEG classification respectively.\\

Based on this combined knowledge of a source's optical and radio luminosities, environment and excitation mode, the results are:
\begin{itemize}\setlength{\itemsep}{0.5mm}
  \item At a given radio luminosity, both accretion modes show similar FR morphological split of sources, overall as well as when restricting the analysis to only rich or only poor environments. This could imply that extended radio morphology is depending only on the power of the resultant jet, and its interactions with the larger-scale environment, and not on the accretion mode of the black hole.
  \item High-excitation galaxies are found almost exclusively in low-density environments while low-excitation galaxies occupy a wider range of densities, independent of FR morphology. This is consistent with the different fuelling mechanisms expected for these excitation modes.
  \item It appears that radio sources in rich clusters have a higher probability of being FRI and show low-excitation. This can be explained by the fact that jets in massive galaxies with low cooling-rates, giving rise to LEGs \citep{Hard07}, are easily disrupted, resulting in FRI-like morphologies in dense environments. On the other hand, a HEG/LEG in a poor/rich environment has roughly equal probabilities of being of morphological type-I or type-II, within errors. However, there is a significant overlap in environment between the two classes, and no clear driving factor between the FRI and FRII sources is found even when combining radio luminosity, accretion mode, large-scale environment and host galaxy luminosity.
    \item The \cite{Led96} relation does not hold for the CoNFIG local sub-sample, even when considering the different intrinsic and extrinsic parameters.
\end{itemize}

The results of this study hint towards the fact that, although originating from two different production mechanisms, the jets of FRI and FRII sources appear to be effectively the same, and to not behave differently in distinct environments. These conclusions are however highly dependent on the errors associated with the samples, in particular on the completeness of HEG/LEG classification (when including these sources based on the idea that they are a random subset of HEGs and LEGs, the Poisson errors become 20\% smaller). Yet, this result is supported by the fact that radio galaxies of different FR type are not hosted by significantly different galaxies, whereas HEGs and LEGs are.\\ 

If intrinsic and large-scale environmental parameters do indeed fail to fully explain the morphological differences between radio sources, it is possible that the distinction FRI/FRII is based on small-scales characteristics, such as the gas mass in the host galaxy (independent of the host mass). This connection between radio morphology and gas mass in the most powerful AGNs in the Local Universe has been previously explored by \cite{Ev05} and \cite{Oca10}. They found that molecular gas mass in FRII is a factor of $\sim$4 greater than in FRI. However, as stated by  \cite{Oca10}, this might be a result of Malmquist bias, with the FRII sources that they study being systematically at higher redshift (and thus showing greater powers) than FRIs. In addition, their samples contained a nearly one-to-one correspondence between FRI and LEG, and between FRII and HEG, meaning that their results could entirely be driven by an underlying LEG/HEG difference in molecular gas properties (as has been established by \cite{Smo12}. To separate out the effects of HEG/LEG and FR differences, and hence understand the causes of jet disruption differences in FRIs and FRIIs, it is essential to investigate cross-populations (FRI HEGs and FRII LEGs), for example using high-resolution ALMA sub-mm observations of sources in the CoNFIG local sub-sample.

\section*{Acknowledgements}
We would like to thank the referee for the helpful comments provided.\\
This work was supported by the National Sciences and Engineering Research Council of Canada (JVW).\\
We thank the staff of the GMRT who have made these observations possible. GMRT is run by the National Centre for Radio Astrophysics of the Tata Institute of Fundamental Research.\\
This work is based in part on data obtained as part of the UKIRT Infrared Deep Sky Survey.\\
This research has made use of the SIMBAD database and the VizieR catalogue access tool, operated at CDS,Strasbourg, France.\\
This research has made use of the NASA/IPAC Extragalactic Database (NED) which is operated by the Jet Propulsion Laboratory, California Institute of Technology, under contract with the National Aeronautics and Space Administration. \\
Funding for the SDSS and SDSS-II has been provided by the Alfred P. Sloan Foundation, the Participating Institutions, the National Science Foundation, the U.S. Department of Energy, the National Aeronautics and Space Administration, the Japanese Monbukagakusho, the Max Planck Society, and the Higher Education Funding Council for England. The SDSS Web Site is http://www.sdss.org/. The SDSS is managed by the Astrophysical Research Consortium for the Participating Institutions. The Participating Institutions are the American Museum of Natural History, Astrophysical Institute Potsdam, University of Basel, University of Cambridge, Case Western Reserve University, University of Chicago, Drexel University, Fermilab, the Institute for Advanced Study, the Japan Participation Group, Johns Hopkins University, the Joint Institute for Nuclear Astrophysics, the Kavli Institute for Particle Astrophysics and Cosmology, the Korean Scientist Group, the Chinese Academy of Sciences (LAMOST), Los Alamos National Laboratory, the Max-Planck-Institute for Astronomy (MPIA), the Max-Planck-Institute for Astrophysics (MPA), New Mexico State University, Ohio State University, University of Pittsburgh, University of Portsmouth, Princeton University, the United States Naval Observatory, and the University of Washington.\\
This publication makes use of data products from the Two Micron All Sky Survey, which is a joint project of the University of Massachusetts and the Infrared Processing and Analysis Center/California Institute of Technology, funded by the National Aeronautics and Space Administration and the National Science Foundation.\\
This research has made use of data obtained from the SuperCOSMOS Science Archive, prepared and hosted by the Wide Field Astronomy Unit, Institute for Astronomy, University of Edinburgh, which is funded by the UK Science and Technology Facilities Council.\\
M.A.G would like to thank the University of Calgary for its hospitality.


\onecolumn
\appendix

\begin{table}
\section{Updated data in the CoNFIG catalogue}
  \begin{center}
    \caption[]{\label{GMRTDat}Flux density measurements at 591-MHz from the GMRT data described in \S\ref{GMRT}.} 
   \medskip
    \begin{tabular}{|cc|cc|cc|cc|cc|}
      \hline
      Source & S$_{591MHz}$ & Source & S$_{591MHz}$ & Source & S$_{591MHz}$ & Source & S$_{591MHz}$ & Source & S$_{591MHz}$ \\ 
             & (mJy) &        & (mJy) &        & (mJy) &        & (mJy) &        & (mJy) \\ 
      \hline
      C1-128 & 2434$\pm$28                          & C4-016 & 113.8$\pm$4.6 & C4-074 & \phantom{1}57.3$\pm$1.9 & C4-116 &           259.2$\pm$4.2 & C4-152 & \phantom{1}25.0$\pm$3.4 \\
      C1-163 & 2585$\pm$63                          & C4-019 & 116.1$\pm$1.9 & C4-084 & \phantom{1}79.7$\pm$2.6 & C4-127 &           120.0$\pm$1.3 & C4-158 & \phantom{1}34.3$\pm$2.2 \\
      C2-095 & 1386$\pm$19                          & C4-021 & 143.0$\pm$2.4 & C4-088 &           134.7$\pm$3.6 & C4-131 & \phantom{1}68.4$\pm$2.0 & C4-166 & \phantom{1}80.1$\pm$4.4 \\
      C3-219 & \phantom{1}219.8$\pm$1.8\phantom{0}  & C4-028 & 115.7$\pm$7.0 & C4-091 & \phantom{1}62.3$\pm$1.0 & C4-132 & \phantom{1}66.7$\pm$2.7 & C4-172 &           119.9$\pm$1.7 \\
      C4-003 & \phantom{1}136.3$\pm$3.9\phantom{0}  & C4-040 & 229.4$\pm$2.0 & C4-093 &           118.0$\pm$2.3 & C4-133 & \phantom{1}50.0$\pm$3.0 & C4-176 & \phantom{1}52.7$\pm$5.4 \\
      C4-004 & \phantom{1}117.4$\pm$2.7\phantom{0}  & C4-044 & 109.2$\pm$9.7 & C4-097 & \phantom{1}87.0$\pm$1.5 & C4-139 & \phantom{1}81.5$\pm$2.7 & C4-178 &           138.7$\pm$7.2 \\
      C4-006 & \phantom{1}149.2$\pm$3.7\phantom{0}  & C4-048 & 135.4$\pm$2.5 & C4-101 & \phantom{1}98.7$\pm$1.9 & C4-140 &           232.4$\pm$2.9 & C4-181 &           210.4$\pm$1.6 \\
      C4-010 & \phantom{1}118.0$\pm$2.3\phantom{0}  & C4-052 & 159.8$\pm$2.1 & C4-102 & \phantom{1}98.3$\pm$2.0 & C4-141 & \phantom{1}54.8$\pm$1.5 & C4-183 &           106.2$\pm$1.9 \\
      C4-011 & \phantom{1}813.3$\pm$5.9\phantom{0}  & C4-056 & 110.1$\pm$1.8 & C4-103 & \phantom{1}78.4$\pm$1.1 & C4-143 & \phantom{1}52.0$\pm$8.3 & & \\
      C4-014 & \phantom{1}128.5$\pm$4.3\phantom{0}  & C4-057 & 113.3$\pm$1.6 & C4-113 &           112.6$\pm$1.3 & C4-144 & \phantom{1}70.1$\pm$2.4 & & \\
      \hline
    \end{tabular}
  \end{center}
\end{table}
\begin{table}
  \begin{center}
    \caption[]{\label{AImprov}Revised/new spectral index values  $\alpha$, defined as $S_{\nu} \propto \nu^{\alpha}, $for sources in
      the CoNFIG catalogue, computed by:\\
    $^v$ including VLSS flux density data\\
    $^c$ including CLASS flux density data\\
    $^g$ including GMRT flux density measurement (as described in \S\ref{GMRT})\\
    $^o$ other\\
    as described in \S\ref{SpecInd}.} 
    \medskip
    \begin{tabular}{|cc||cc||cc||cc||cc||cc|}
      \hline
      Source & $\alpha$ & Source & $\alpha$ & Source & $\alpha$ &
      Source & $\alpha$ & Source & $\alpha$ & Source & $\alpha$ \\ 
      \hline
      C1-001 & $-$0.27$^c\phantom{v}$   & C2-228 & $-$0.81$^v\phantom{v}$   & C3-195 & $-$1.04$^c\phantom{v}$ & C4-035 & $-$0.84$^v\phantom{v}$ & C4-093 & $-$0.60$^g\phantom{v}$ & C4-152 & \phantom{$-$}1.07$^g\phantom{v}$ \\
      C1-002 & $-$0.33$^c\phantom{v}$   & C3-002 & $-$0.59$^v\phantom{v}$   & C3-201 & $-$0.73$^v\phantom{v}$ & C4-040 & $-$0.79$^g\phantom{v}$ & C4-097 & $-$0.31$^g\phantom{v}$ & C4-153 & $-$0.92$^v\phantom{v}$ \\
      C1-076 & $-$0.30$^c\phantom{v}$   & C3-006 & $-$0.58$^o\phantom{v}$   & C3-206 & $-$0.60$^v\phantom{v}$ & C4-041 & $-$0.93$^v\phantom{v}$ & C4-098 & $-$0.69$^v\phantom{v}$ & C4-157 & $-$0.10$^o\phantom{v}$ \\
      C1-138 & $-$0.47$^v\phantom{v}$   & C3-010 & $-$0.27$^c\phantom{v}$   & C3-216 & $-$0.57$^v\phantom{v}$ & C4-044 & $-$0.79$^g\phantom{v}$ & C4-101 & $-$0.14$^g\phantom{v}$ & C4-158 & \phantom{$-$}0.45$^g\phantom{v}$ \\
      C1-175 & $-$0.35$^o\phantom{v}$   & C3-012 & $-$0.82$^v\phantom{v}$   & C3-219 & $-$0.11$^g\phantom{v}$ & C4-047 & $-$0.62$^o\phantom{v}$ & C4-102 & $-$0.61$^g\phantom{v}$ & C4-163 & $-$0.81$^v\phantom{v}$ \\
      C1-181 & $-$0.19$^o\phantom{v}$   & C3-018 & $-$0.37$^c\phantom{v}$   & C3-244 & $-$0.62$^v\phantom{v}$ & C4-048 & $-$0.78$^g\phantom{v}$ & C4-103 & $-$0.30$^g\phantom{v}$ & C4-166 & $-$0.21$^g\phantom{v}$ \\
      C1-198 & $-$0.58$^c\phantom{v}$   & C3-024 & $-$0.06$^c\phantom{v}$   & C3-274 & $-$0.66$^v\phantom{v}$ & C4-049 & $-$0.95$^v\phantom{v}$ & C4-107 & $-$0.98$^v\phantom{v}$ & C4-167 & $-$0.14$^o\phantom{v}$ \\
      C1-215 & $-$0.42$^{cv}$ & C3-025 & $-$0.48$^v\phantom{v}$   & C3-280 & $-$0.71$^v\phantom{v}$ & C4-050 & $-$0.46$^o\phantom{v}$ & C4-113 & $-$0.72$^g\phantom{v}$ & C4-168 & $-$0.91$^v\phantom{v}$ \\
      C1-233 & $-$0.32$^{cv}$ & C3-027 & $-$0.62$^v\phantom{v}$   & C3-281 & $-$0.67$^v\phantom{v}$ & C4-052 & $-$0.71$^g\phantom{v}$ & C4-116 & $-$1.09$^g\phantom{v}$ & C4-169 & $-$1.54$^c\phantom{v}$ \\
      C1-236 & $-$0.90$^c\phantom{v}$   & C3-044 & $-$0.54$^v\phantom{v}$   & C3-286 & $-$0.58$^c\phantom{v}$ & C4-055 & $-$0.74$^o\phantom{v}$ & C4-120 & $-$0.74$^v\phantom{v}$ & C4-172 & $-$0.54$^g\phantom{v}$ \\
      C1-239 & $-$0.36$^{cv}$ & C3-047 & $-$0.36$^c\phantom{v}$   & C4-003 & $-$0.93$^g\phantom{v}$ & C4-056 & $-$0.19$^g\phantom{v}$ & C4-127 & $-$0.81$^g\phantom{v}$ & C4-173 & $-$0.47$^o\phantom{v}$ \\
      C2-009 & $-$0.37$^c\phantom{v}$   & C3-051 & $-$0.06$^c\phantom{v}$   & C4-004 & $-$0.54$^g\phantom{v}$ & C4-057 & $-$0.72$^g\phantom{v}$ & C4-128 & $-$0.86$^v\phantom{v}$ & C4-174 & $-$1.41$^c\phantom{v}$ \\
      C2-032 & $-$0.43$^c\phantom{v}$   & C3-063 & $-$0.55$^v\phantom{v}$   & C4-006 & $-$0.76$^g\phantom{v}$ & C4-066 & $-$0.21$^c\phantom{v}$ & C4-131 & $-$0.23$^g\phantom{v}$ & C4-176 & \phantom{$-$}0.09$^g\phantom{v}$ \\
      C2-059 & $-$0.44$^c\phantom{v}$   & C3-069 & $-$0.72$^v\phantom{v}$   & C4-008 & $-$1.00$^v\phantom{v}$ & C4-067 & $-$0.71$^v\phantom{v}$ & C4-132 & $-$0.15$^g\phantom{v}$ & C4-178 & $-$0.38$^g\phantom{v}$ \\
      C2-062 & $-$0.27$^o\phantom{v}$   & C3-078 & $-$0.45$^{cv}$ & C4-010 & $-$0.62$^g\phantom{v}$ & C4-071 & $-$1.53$^o\phantom{v}$ & C4-133 & \phantom{$-$}0.18$^g\phantom{v}$ & C4-180 & $-$0.39$^o\phantom{v}$ \\
      C2-102 & $-$0.61$^c\phantom{v}$   & C3-079 & $-$0.47$^o\phantom{v}$   & C4-011 & \phantom{$-$}0.13$^g\phantom{v}$ & C4-072 & $-$0.87$^v\phantom{v}$ & C4-134 & $-$0.79$^v\phantom{v}$ & C4-181 & \phantom{$-$}0.03$^g\phantom{v}$ \\
      C2-112 & $-$0.23$^{cv}$ & C3-094 & $-$0.77$^v\phantom{v}$   & C4-014 & \phantom{$-$}0.28$^g\phantom{v}$ & C4-074 & $-$0.07$^g\phantom{v}$ & C4-135 & $-$0.78$^v\phantom{v}$ & C4-183 & $-$0.31$^g\phantom{v}$ \\
      C2-155 & $-$0.56$^{cv}$ & C3-116 & $-$1.12$^c\phantom{v}$   & C4-015 & $-$0.78$^o\phantom{v}$ & C4-078 & $-$0.77$^v\phantom{v}$ & C4-139 & $-$0.17$^g\phantom{v}$ & C4-184 & $-$0.75$^v\phantom{v}$ \\
      C2-161 & $-$0.27$^c\phantom{v}$   & C3-123 & $-$0.88$^c\phantom{v}$   & C4-016 & $-$0.54$^g\phantom{v}$ & C4-080 & $-$0.74$^v\phantom{v}$ & C4-140 & $-$0.16$^g\phantom{v}$ & C4-185 & $-$0.85$^v\phantom{v}$ \\
      C2-162 & $-$0.11$^c\phantom{v}$   & C3-137 & $-$1.02$^o\phantom{v}$   & C4-019 & $-$0.89$^g\phantom{v}$ & C4-082 & $-$0.88$^v\phantom{v}$ & C4-141 & \phantom{$-$}1.78$^g\phantom{v}$ & \phantom{$-$}    & \phantom{$-$}     \\
      C2-165 & $-$0.86$^c\phantom{v}$   & C3-139 & $-$0.49$^v\phantom{v}$   & C4-021 & $-$1.13$^g\phantom{v}$ & C4-084 & $-$0.51$^g\phantom{v}$ & C4-142 & $-$0.66$^v\phantom{v}$ &        &         \\
      C2-173 & $-$0.82$^c\phantom{v}$   & C3-146 & $-$0.63$^v\phantom{v}$   & C4-022 & $-$1.17$^v\phantom{v}$ & C4-085 & $-$0.24$^o\phantom{v}$ & C4-143 & \phantom{$-$}0.03$^g\phantom{v}$ &        &         \\
      C2-193 & $-$0.18$^c\phantom{v}$   & C3-173 & $-$0.51$^{cv}$ & C4-025 & $-$0.50$^{cv}$ & C4-088 & $-$0.81$^g\phantom{v}$ & C4-144 & $-$0.31$^g\phantom{v}$ &        &         \\
      C2-200 & $-$0.84$^o\phantom{v}$   & C3-181 & $-$0.72$^v\phantom{v}$   & C4-028 & $-$0.34$^g\phantom{v}$ & C4-091 & $-$0.19$^g\phantom{v}$ & C4-146 & $-$0.63$^{cv}$ &        &         \\    
      \hline
    \end{tabular}
  \end{center}
\end{table}

\begin{table}
  \begin{center}
    \caption[]{\label{UKIDSS}UKIDSS K-band magnitudes for sources in the CoNFIG catalogue, as defined in \S\ref{ZandK}, with a 2.0 arcsec aperture diameter. Note that a minimum error of $\Delta K$=0.1 is assigned.\\
      $^n$ indicates sources for which this is the first detection of the host-galaxy.}
    \medskip
    \begin{tabular}{llllllll}
      \hline
      Source & \multicolumn{1}{c}{K} & Source & \multicolumn{1}{c}{K} & Source & \multicolumn{1}{c}{K} & Source & \multicolumn{1}{c}{K} \\
      \hline
      C1-011    & 13.6$\pm$0.1 & C1-247    & 15.6$\pm$0.1 & C3-122    & 18.1$\pm$0.1 & C4-073$^n$& 18.2$\pm$0.2 \\
      C1-018    & 14.4$\pm$0.1 & C1-272    & 15.3$\pm$0.1 & C3-127    & 17.4$\pm$0.1 & C4-074    & 16.4$\pm$0.1 \\
      C1-021    & 15.7$\pm$0.1 & C2-010    & 16.0$\pm$0.1 & C3-134    & 15.7$\pm$0.1 & C4-079$^n$& 18.0$\pm$0.1 \\
      C1-036    & 18.0$\pm$0.2 & C2-012$^n$& 18.0$\pm$0.1 & C3-142    & 13.4$\pm$0.1 & C4-080    & 16.8$\pm$0.1 \\
      C1-038    & 15.3$\pm$0.1 & C2-014    & 15.9$\pm$0.1 & C3-144    & 17.1$\pm$0.1 & C4-081    & 17.2$\pm$0.1 \\
      C1-054    & 16.8$\pm$0.1 & C2-019    & 15.9$\pm$0.1 & C3-153    & 17.1$\pm$0.1 & C4-085    & 14.4$\pm$0.1 \\
      C1-055    & 17.0$\pm$0.1 & C2-035    & 17.9$\pm$0.1 & C3-159$^n$& 18.5$\pm$0.2 & C4-086    & 17.2$\pm$0.1 \\
      C1-056    & 12.6$\pm$0.1 & C2-036    & 16.2$\pm$0.1 & C3-166    & 15.7$\pm$0.1 & C4-088    & 17.8$\pm$0.1 \\
      C1-059    & 13.4$\pm$0.1 & C2-038    & 18.1$\pm$0.1 & C3-167    & 15.9$\pm$0.1 & C4-092    & 16.2$\pm$0.1 \\
      C1-066    & 17.4$\pm$0.1 & C2-046    & 16.3$\pm$0.1 & C3-180    & 15.0$\pm$0.1 & C4-093    & 15.4$\pm$0.1 \\
      C1-077    & 15.2$\pm$0.1 & C2-049    & 14.6$\pm$0.1 & C3-189    & 13.1$\pm$0.1 & C4-094    & 16.7$\pm$0.1 \\
      C1-078    & 16.7$\pm$0.1 & C2-052    & 16.0$\pm$0.1 & C3-195    & 13.6$\pm$0.1 & C4-097$^n$& 17.9$\pm$0.2 \\
      C1-082    & 13.2$\pm$0.1 & C2-062    & 15.6$\pm$0.1 & C3-199$^n$& 18.1$\pm$0.1 & C4-098    & 14.8$\pm$0.1 \\
      C1-104    & 16.4$\pm$0.1 & C2-065    & 14.8$\pm$0.1 & C3-208    & 14.9$\pm$0.1 & C4-101    & 17.0$\pm$0.1 \\
      C1-111    & 14.8$\pm$0.1 & C2-069    & 17.2$\pm$0.1 & C3-246    & 17.0$\pm$0.1 & C4-107    & 16.5$\pm$0.1 \\
      C1-121    & 17.0$\pm$0.1 & C2-085    & 17.6$\pm$0.1 & C3-250$^n$& 16.6$\pm$0.1 & C4-111    & 15.4$\pm$0.1 \\
      C1-128    & 13.6$\pm$0.1 & C2-094    & 13.7$\pm$0.1 & C3-253    & 17.2$\pm$0.1 & C4-115    & 15.5$\pm$0.1 \\
      C1-129    & 11.6$\pm$0.1 & C2-103$^n$& 17.5$\pm$0.1 & C4-001    & 17.5$\pm$0.1 & C4-118    & 15.9$\pm$0.1 \\
      C1-133    & 11.1$\pm$0.1 & C2-117    & 15.2$\pm$0.1 & C4-002    & 15.8$\pm$0.1 & C4-119$^n$& 17.1$\pm$0.1 \\
      C1-135    & 10.2$\pm$0.1 & C2-123    & 15.6$\pm$0.1 & C4-003    & 17.6$\pm$0.1 & C4-122$^n$& 17.7$\pm$0.1 \\
      C1-136    & 13.5$\pm$0.1 & C2-126    & 14.5$\pm$0.1 & C4-004$^n$& 18.3$\pm$0.2 & C4-123    & 17.9$\pm$0.1 \\
      C1-144    & 15.8$\pm$0.1 & C2-131    & 17.9$\pm$0.1 & C4-005$^n$& 17.7$\pm$0.1 & C4-125    & 17.8$\pm$0.1 \\
      C1-147    & 17.6$\pm$0.1 & C2-133    & 14.9$\pm$0.1 & C4-007$^n$& 18.1$\pm$0.2 & C4-126    & 17.4$\pm$0.1 \\
      C1-152    & 18.3$\pm$0.2 & C2-153    & 15.0$\pm$0.1 & C4-008    & 18.2$\pm$0.2 & C4-131    & 16.9$\pm$0.1 \\
      C1-153    & 16.2$\pm$0.1 & C2-171    & 16.1$\pm$0.1 & C4-016    & 15.6$\pm$0.1 & C4-137    & 16.2$\pm$0.1 \\
      C1-159    & 15.4$\pm$0.1 & C2-188    & 15.7$\pm$0.1 & C4-020    & 14.8$\pm$0.1 & C4-139$^n$& 18.1$\pm$0.2 \\
      C1-161    & 16.0$\pm$0.1 & C2-191    & 15.3$\pm$0.1 & C4-023$^n$& 18.0$\pm$0.1 & C4-142    & 15.8$\pm$0.1 \\
      C1-168    & 14.4$\pm$0.1 & C2-193    & 16.3$\pm$0.1 & C4-025    & 16.4$\pm$0.1 & C4-145    & 16.6$\pm$0.1 \\
      C1-175    & 14.2$\pm$0.1 & C2-196    & 16.4$\pm$0.1 & C4-027$^n$& 18.0$\pm$0.1 & C4-146    & 14.0$\pm$0.1 \\
      C1-177    & 17.2$\pm$0.1 & C2-204    & 16.7$\pm$0.1 & C4-028    & 15.3$\pm$0.1 & C4-153    & 17.5$\pm$0.1 \\
      C1-178    & 17.1$\pm$0.1 & C2-208    & 16.3$\pm$0.1 & C4-029    & 15.4$\pm$0.1 & C4-155    & 16.9$\pm$0.1 \\
      C1-180    & 17.4$\pm$0.1 & C2-220    & 13.8$\pm$0.1 & C4-035    & 17.6$\pm$0.1 & C4-156$^n$& 17.9$\pm$0.1 \\
      C1-193    & 17.8$\pm$0.1 & C2-233    & 14.1$\pm$0.1 & C4-039    & 17.9$\pm$0.1 & C4-159    & 17.2$\pm$0.1 \\
      C1-194    & 13.0$\pm$0.1 & C2-239    & 14.5$\pm$0.1 & C4-042    & 17.5$\pm$0.1 & C4-161    & 16.3$\pm$0.1 \\
      C1-198    & 15.7$\pm$0.1 & C3-001    & 17.7$\pm$0.1 & C4-043    & 16.2$\pm$0.1 & C4-166    & 14.5$\pm$0.1 \\
      C1-199    & 16.0$\pm$0.1 & C3-006    & 17.3$\pm$0.1 & C4-044    & 14.9$\pm$0.1 & C4-169    & 17.8$\pm$0.2 \\
      C1-204    & 15.2$\pm$0.1 & C3-016    & 16.1$\pm$0.1 & C4-049    & 14.2$\pm$0.1 & C4-170    & 18.0$\pm$0.2 \\
      C1-207    & 15.9$\pm$0.1 & C3-022    & 15.5$\pm$0.1 & C4-050    & 14.7$\pm$0.1 & C4-172    & 16.1$\pm$0.1 \\
      C1-208    & 17.3$\pm$0.1 & C3-047    & 16.4$\pm$0.1 & C4-051$^n$& 18.1$\pm$0.2 & C4-173    & 18.4$\pm$0.2 \\
      C1-211    & 12.7$\pm$0.1 & C3-049    & 16.3$\pm$0.1 & C4-052    & 17.3$\pm$0.1 & C4-174    & 16.5$\pm$0.1 \\
      C1-215    & 17.7$\pm$0.1 & C3-057    & 12.3$\pm$0.1 & C4-054    & 17.2$\pm$0.1 & C4-176    & 15.5$\pm$0.1 \\
      C1-220    & 15.9$\pm$0.1 & C3-060$^n$& 18.1$\pm$0.1 & C4-055    & 14.0$\pm$0.1 & C4-178    & 17.2$\pm$0.1 \\
      C1-225    & 16.2$\pm$0.1 & C3-070    & 18.3$\pm$0.2 & C4-062    & 17.1$\pm$0.1 & C4-179    & 17.5$\pm$0.1 \\
      C1-229    & 14.1$\pm$0.1 & C3-093    & 14.6$\pm$0.1 & C4-064    & 17.6$\pm$0.1 & C4-180    & 16.9$\pm$0.1 \\
      C1-236    & 16.6$\pm$0.1 & C3-097    & 17.5$\pm$0.1 & C4-066    & 17.9$\pm$0.1 & C4-184    & 15.1$\pm$0.1 \\
      C1-238    & 16.8$\pm$0.1 & C3-101    & 17.9$\pm$0.1 & C4-068$^n$& 18.2$\pm$0.2 & C4-188    & 17.5$\pm$0.1 \\
      C1-240    & 18.0$\pm$0.1 & C3-104    & 13.4$\pm$0.1 & C4-071    & 16.2$\pm$0.1 &           &              \\
      C1-245    & 15.4$\pm$0.1 & C3-105    & 14.7$\pm$0.1 & C4-072    & 16.7$\pm$0.1 &           &              \\
      \hline
    \end{tabular}
  \end{center}
\end{table}
\begin{table}
  \begin{center}
    \caption[]{\label{UKIDSScoord}Coordinates of the 20 new UKIDSS optical identifications for sources in the CoNFIG catalogue, as defined in \S\ref{ZandK}.}
    \medskip
    \begin{tabular}{llll}
      \hline
      Source & \multicolumn{1}{c}{Coord. (J2000)} & Source & \multicolumn{1}{c}{Coord. (J2000)} \\
      \hline
      C2-012 & 09 36 31.97   +04 22 10.02 & C4-027 & 14 11 10.29 $-$00 36 01.67\\
      C2-103 & 11 11 22.64   +03 09 09.67 & C4-051 & 14 15 30.52   +02 23 02.50\\
      C3-060 & 14 56 28.71   +13 02 40.58 & C4-068 & 14 19 13.52 $-$00 13 51.21\\
      C3-159 & 15 18 35.95   +10 32 12.26 & C4-073 & 14 20 34.15 $-$00 54 59.92\\
      C3-199 & 15 31 47.96   +10 55 33.20 & C4-079 & 14 23 03.45   +01 39 58.50\\
      C3-250 & 15 50 11.83   +27 17 59.40 & C4-097 & 14 26 12.95   +02 00 39.38\\
      C4-004 & 14 08 32.70 $-$01 31 20.78 & C4-119 & 14 30 00.91   +00 46 26.51\\
      C4-005 & 14 08 33.36   +01 16 22.05 & C4-122 & 14 30 30.63   +01 01 03.14\\
      C4-007 & 14 08 46.80   +01 33 56.27 & C4-139 & 14 33 08.85   +00 44 34.90\\
      C4-023 & 14 10 35.35 $-$00 41 53.03 & C4-156 & 14 36 30.35   +00 35 19.05\\
      \hline
    \end{tabular}
  \end{center}
\end{table}
\begin{table}
  \begin{center}
    \caption[]{\label{ZImprov}Revised redshift for sources in the CoNFIG catalogue. \\
References: (1) SDSS spectroscopic redshift; (2)
      \cite{Tinti06}; (3) \cite{Rich09}; (4) \cite{White92}; (5) \cite{Croom09}; (6) UKIDSS K-z relation}
    \medskip
    \begin{tabular}{cccccccc}
      \hline
      Source & Redshift & Source & Redshift & Source & Redshift & Source & Redshift \\ 
      \hline
      C1-034 & 4.5165$^1$ & C3-027 & 2.2550$^3$ & C3-168 & 3.2253$^1$ & C4-043 & 0.4000$^6$ \\
      C1-062 & 0.8993$^1$ & C3-032 & 0.2505$^1$ & C3-171 & 2.1824$^1$ & C4-051 & 1.6000$^6$ \\
      C1-082 & 0.3823$^1$ & C3-048 & 0.3350$^3$ & C3-188 & 2.7950$^3$ & C4-068 & 1.8000$^6$ \\
      C1-086 & 0.5500$^2$ & C3-051 & 1.2760$^1$ & C3-194 & 2.2650$^3$ & C4-081 & 0.8000$^6$ \\
      C1-185 & 0.2600$^4$ & C3-060 & 1.6000$^6$ & C3-199 & 1.6000$^6$ & C4-088 & 1.3000$^6$ \\
      C1-213 & 0.5798$^1$ & C3-070 & 1.8000$^6$ & C3-222 & 2.5424$^1$ & C4-092 & 0.4000$^6$ \\
      C2-038 & 1.6000$^6$ & C3-071 & 1.6850$^3$ & C3-250 & 0.6000$^6$ & C4-119 & 0.8000$^6$ \\
      C2-085 & 0.6500$^2$ & C3-099 & 2.2830$^1$ & C4-005 & 1.2000$^6$ & C4-122 & 1.2000$^6$ \\
      C2-103 & 1.1000$^6$ & C3-101 & 1.5450$^3$ & C4-007 & 1.6000$^6$ & C4-125 & 1.3000$^6$ \\
      C2-185 & 0.6793$^1$ & C3-108 & 1.8247$^1$ & C4-008 & 1.7000$^6$ & C4-153 & 1.0000$^6$ \\
      C2-233 & 0.3183$^1$ & C3-122 & 1.6000$^6$ & C4-013 & 1.6250$^3$ & C4-155 & 0.2250$^3$ \\
      C3-003 & 0.9920$^1$ & C3-132 & 1.6631$^1$ & C4-023 & 1.5000$^6$ & C4-159 & 0.9632$^5$ \\
      C3-018 & 1.5550$^3$ & C3-144 & 0.8000$^6$ & C4-027 & 1.5000$^6$ & C4-170 & 1.4000$^6$ \\
      C3-024 & 1.0236$^1$ & C3-147 & 0.5798$^1$ & C4-039 & 1.4000$^6$ &        &            \\
      \hline
    \end{tabular}
  \end{center}
\end{table}

\begin{landscape}
  \begin{table}
    \section{CoNFIG local sub-group}\label{LocalTab} 
    \caption[]{\label{Local} Spectral features, richness factor and I-band magnitude of local (z $\le$ 0.3) sources in the CoNFIG catalogue.}
    Flux and rest-frame equivalent width (EW) of the lines are given in units of $\AA$ and $10^{-17} erg/cm^2/s/\AA$ respectively. Details of the HEG/LEG classification can be found in \S\ref{Excitation}. The richness factors are defined in section \S\ref{Env}.\\
    Column 5 (M) specify the source morphology. I - FRI; II - FRII; U - Unclassified extended; C - Compact\\
    Column 15 (Band) specify which catalogue \& optical band was used to determine the richness factor. r - SDSS r-magnitude; R - SSS R-magnitude; I - EIS I magnitude.\\
    Values of I in column 16 with a $^*$ were derived from SDSS {\it i}-band magnitude values.\\
    Spectrum reference: 1 - SDSS; 2 - 3CRR emission line catalogue (https://www.astrosci.ca/users/willottc/3crr/3crr.html); 3 - \cite{Butti09}; 4 - \cite{Ho95}; 5 - \cite{White00}; 6 - 2dFGRS; 7 - \cite{Brookes07}\\\\
    \centerline{
    \begin{tabular}{llrrl|rrrrcc|rc|c}
      \hline
      \multicolumn{1}{c}{ID} & \multicolumn{1}{c}{Name} & \multicolumn{1}{c}{RA} & \multicolumn{1}{c}{DEC} & M. & \multicolumn{2}{c}{$\rm [OII]_{3727\AA}$} & \multicolumn{2}{c}{$\rm [OIII]_{5007\AA}$}  & HEG/ & \multicolumn{1}{c}{spec.} & \multicolumn{1}{c}{Richness} & Band & I-mag \\
      & & \multicolumn{2}{c}{(J2000)}    & & \multicolumn{1}{c}{flux} & \multicolumn{1}{c}{EW} & \multicolumn{1}{c}{flux} & \multicolumn{1}{c}{EW} & LEG  & \multicolumn{1}{c}{ref}& & & \\
        \hline
3C     & 3C31        & 01 07 24.95 & +32 24 45.15   & I  &           &         &         &          & L & 2 &    183.32 & R &   5.75 \\
3C     & 3C33        & 01 08 52.86 & +13 20 14.36   & II &           &         &         &          & H & 2 &     18.71 & R &  15.71 \\
3C     & 3C33.1      & 01 09 44.27 & +73 11 57.33   & II &    45.628 &  28.507 & 215.154 &  263.425 & H & 3 &      4.81 & R &  19.31 \\
3C     & 3C61.1      & 02 22 35.18 & +86 19 06.51   & II &           &         &         &          & H & 2 &     15.97 & R &  19.21 \\
3C     & 3C66B       & 02 23 11.41 & +42 59 31.51   & I  &           &         &         &          & L & 2 & $-$263.07 & R &  16.75 \\
3C     & 3C79        & 03 10 00.08 & +17 05 58.65   & II &           &         &         &          & H & 2 &     -3.97 & R &  17.18 \\
3C     & 3C83.1B     & 03 18 15.69 & +41 52 27.99   & I  &           &         &         &          & L & 3 &    148.39 & R &  11.47 \\
3C     & 3C84        & 03 19 48.14 & +41 30 42.35   & I  &           &         &         &          & L & 2 &    114.53 & R &   6.49 \\
3C     & 3C98        & 03 58 54.43 & +10 26 02.81   & II &           &         &         &          & H & 2 &  $-$18.65 & R &  14.65 \\
3C     & 3C123       & 04 37 04.37 & +29 40 13.86   & II &           &         &         &          & L & 2 &     15.26 & R &  18.32 \\
3C     & 3C133       & 05 02 58.50 & +25 16 24.00   & II &           &         &         &          & H & 2 &      7.57 & R &  19.47 \\
3C     & 3C153       & 06 09 32.53 & +48 04 15.35   & II &           &         &         &          & L & 2 &      9.67 & R &  16.54 \\
3C     & 3C171       & 06 55 14.73 & +54 08 57.39   & II &           &         &         &          & H & 2 &  $-$41.22 & R &  17.16 \\
3C     & 3C231       & 09 55 52.92 & +69 40 46.14   & I  &           &         &         &          & L & 4 & $-$247.57 & R &   7.40 \\
3C     & 3C382       & 18 35 03.37 & +32 41 46.93   & II &           &         &         &          & H & 3 &    264.15 & R &  14.21 \\
3C     & 3C386       & 18 38 26.22 & +17 11 50.16   & I  &           &         &         &          & L & 3 &    164.82 & R &  13.67 \\
3C     & 3C388       & 18 44 02.35 & +45 33 29.55   & II &           &         &         &          & H & 2 &     73.24 & R &  14.21 \\
3C     & 3C390.3     & 18 42 08.92 & +79 46 17.20   & II &           &         &         &          & H & 3 &     10.68 & R &  15.42 \\
3C     & 3C401       & 19 40 25.01 & +60 41 36.14   & II &           &         &         &          & L & 2 &     38.79 & R &  16.60 \\
3C     & 3C433       & 21 23 44.55 & +25 04 28.04   & II &           &         &         &          & H & 2 &     22.03 & R &  16.47 \\
3C     & 3C438       & 21 55 52.25 & +38 00 28.46   & II &           &         &         &          & H & 2 &     51.99 & R &  17.81 \\
3C     & 3C452       & 22 45 48.75 & +39 41 15.89   & II &           &         &         &          & H & 2 &     65.29 & R &  15.38 \\
3C     & 3C465       & 23 38 29.39 & +27 01 53.53   & I  &           &         &         &          & L & 2 &     72.87 & R &   6.02 \\
C1-003 & 4C 53.16    & 07 16 41.09 & +53 23 10.30   & II &           &         &         &          &   &   &     73.24 & R &  13.29 \\
C1-007 & DA 240      & 07 49 48.10 & +55 54 21.00   & II &           &         &         &          & L & 1 &     14.34 & R &  18.68 \\
C1-008 & NGC 2484    & 07 58 28.60 & +37 47 13.80   & I  &    19.609 &   1.366 &  18.528 &    1.638 & L & 1 &     29.82 & r &  12.22 \\
C1-011 & 3C 192      & 08 05 31.31 & +24 10 21.30   & II &   121.078 &  32.772 & 535.254 &   58.885 & H & 1 &     60.14 & r &  15.15 \\
C1-015 & 4C 52.18    & 08 19 47.55 & +52 32 29.50   & II &           &         &         &          &   &   &     49.28 & r &  17.79 \\
C1-016 & 3C 197.1    & 08 21 33.77 & +47 02 35.70   & II &     0.940 &   1.363 &  18.116 &    9.908 & H & 1 &     15.06 & r &  16.26 \\
C1-017 & 4C 17.44    & 08 21 44.02 & +17 48 20.50   & C  &     1.504 &   4.444 &   4.552 &   12.601 & H & 1 &   $-$0.00 & r &  17.58 \\
C1-025 & 4C 55.16    & 08 34 54.91 & +55 34 21.00   & C  &    65.541 & 218.714 &  42.670 &  125.314 & H & 1 &     40.90 & r &  16.21 \\
C1-026 & 4C 45.17    & 08 37 53.51 & +44 50 54.60   & II &     4.688 &   6.037 &  88.094 &   56.810 & H & 1 &   $-$5.46 & r &  16.44 \\
C1-030 & NGC 2656    & 08 47 53.83 & +53 52 36.80   & I  &           &         &         &          &   &   &     39.47 & r &  15.15 \\
C1-031 & 4C 31.32    & 08 47 57.00 & +31 48 40.50   & II &     6.363 &   1.049 &   3.764 &    0.148 & L & 1 &      4.77 & r &  13.28 \\
      \hline
    \end{tabular}}
\end{table}
\end{landscape}
\begin{landscape}
  \begin{table}
    {\bf Table~\ref{Local} cont.\\}
    \centerline{
    \begin{tabular}{llrrl|rrrrcc|rc|c}
      \hline
      \multicolumn{1}{c}{ID} & \multicolumn{1}{c}{Name} & \multicolumn{1}{c}{RA} & \multicolumn{1}{c}{DEC} & M. & \multicolumn{2}{c}{$\rm [OII]_{3727\AA}$} & \multicolumn{2}{c}{$\rm [OIII]_{5007\AA}$}  & HEG/ & \multicolumn{1}{c}{spec.} & \multicolumn{1}{c}{Richness} & Band & I-mag \\
      & & \multicolumn{2}{c}{(J2000)}    & & \multicolumn{1}{c}{flux} & \multicolumn{1}{c}{EW} & \multicolumn{1}{c}{flux} & \multicolumn{1}{c}{EW} & LEG  & \multicolumn{1}{c}{ref}& & & \\
        \hline 
C1-038 & 3C 213.1    & 09 01 05.40 & +29 01 45.70   & II &    14.247 &  35.530 &  10.456 &  14.070 & L & 1&     13.37 & r &  17.06 \\
C1-046 & 3C 219      & 09 21 07.54 & +45 38 45.70   & II &     6.088 &   6.001 &  80.084 &  56.496 & H & 1&     17.20 & r &  16.16 \\
C1-050 & 3C 223      & 09 39 50.20 & +35 55 53.10   & II &    56.397 &  63.187 & 441.164 & 314.377 & H & 1&     28.41 & r &  16.40 \\
C1-051 & 3C 223.1    & 09 41 23.62 & +39 44 14.10   & II &    24.687 &   9.021 & 211.159 &  54.628 & H & 1&     33.17 & r &  15.30 \\
C1-056 & 3C 227      & 09 47 47.27 & +07 25 13.80   & II &  $-$8.599 &   2.204 & 371.339 &  76.701 & H & 1&     44.76 & r &  15.95 \\
C1-063 & 3C 234      & 10 01 46.73 & +28 46 56.50   & II &    96.280 &  40.353 &1419.283 & 717.561 & H & 1&     18.41 & r &  16.14 \\
C1-064 & 3C 236      & 10 06 01.74 & +34 54 10.40   & II &    14.244 &  11.762 &  22.169 &  12.268 & H & 1&     10.19 & r &  15.04 \\
C1-069 & 4C 39.29    & 10 17 14.15 & +39 01 24.00   & II &           &         &         &         & L & 1&     85.05 & r &  18.91$^*$ \\
C1-070 & 4C 48.29A   & 10 20 49.61 & +48 32 04.20   & II &    29.706 &  12.696 &   7.682 &   1.425 & L & 1&  $-$11.25 & r &  15.63 \\
C1-072 & 4C 59.13    & 10 23 38.71 & +59 04 49.50   & II &           &         &         &         &   &  &   $-$3.23 & r &  18.93 \\
C1-090 & 3C 253      & 11 13 32.13 & $-$02 12 55.20 & II &           &         &         &         &   &  &     27.98 & R &  19.31 \\
C1-092 & 4C 29.41    & 11 16 34.70 & +29 15 20.50   & I  &           &         &         &         &   &  &     61.50 & r &  13.89 \\
C1-101 & 4C 61.23    & 11 37 16.95 & +61 20 38.40   & II &   136.718 &  96.028 & 577.092 & 280.689 & H & 1&     14.88 & r &  16.16 \\
C1-102 & 4C 12.42    & 11 40 27.69 & +12 03 07.60   & I  &    13.929 &   4.485 &   7.669 &   1.137 & L & 1&  $-$21.84 & r &  14.65 \\
C1-106 & 4C 37.32    & 11 44 34.45 & +37 10 16.90   & II &  $-$3.230 &$-$0.315 &  26.704 &   5.616 & H & 1&     17.64 & r &  15.88 \\
C1-107 & 3C 264      & 11 45 05.23 & +19 36 37.80   & I  &    46.097 &   1.626 &  10.391 &   0.195 & L & 1&     90.49 & r &   6.54 \\
C1-114 & 4C 55.22    & 11 55 26.63 & +54 54 13.60   & II &     6.559 &   0.981 &   0.312 &   0.002 & L & 1&   $-$2.25 & r &  13.66 \\
C1-115 & 4C 59.17    & 11 56 03.67 & +58 47 05.40   & U  &           &         &         &         &   &  &      2.41 & r &  18.07 \\
C1-120 & 4C $-$04.40 & 12 04 02.13 & $-$04 22 43.90 & II &           &         &         &         &   &  &      6.56 & R &  15.72 \\
C1-128 & 4C 04.41    & 12 17 29.83 & +03 36 44.00   & I  &  $-$7.594 &$-$8.168 &$-$0.698 &$-$2.488 & L & 1&     80.10 & r &  14.43 \\
C1-129 & 3C 270      & 12 19 15.33 & +05 49 40.40   & I  &           &         &$-$10.643&$-$1.581 & L & 1&     57.51 & r &  10.37 \\
C1-133 & M84         & 12 25 03.78 & +12 52 35.20   & I  &           &         &$-$1.644 &$-$2.737 & L & 1&     93.11 & r &   8.69$^*$ \\
C1-135 & 3C 273      & 12 29 06.41 & +02 03 05.10   & C  &           &         &         &         & H & 3&  $-$11.57 & r &  11.84$^*$ \\
C1-136 & 1227+119    & 12 29 51.84 & +11 40 24.20   & I  &     2.589 &   0.475 & $-$1.233& $-$3.123& L & 1&     93.44 & r &  14.07 \\
C1-137 & M87         & 12 30 49.46 & +12 23 21.60   & I  &           &         &   0.568 &   0.001 & L & 1&    132.82 & r &   9.98 \\
C1-140 & 4C 16.33    & 12 36 29.13 & +16 32 32.10   & I  &           &         &         &         &   &  &     57.01 & r &  14.40 \\
C1-144 & 4C 09.44    & 12 51 44.47 & +08 56 27.80   & II &           &         &         &         &   &  &     91.09 & r &  17.84 \\
C1-146 & 4C 02.34    & 12 53 03.55 & +02 38 22.30   & II &           &         &         &         &   &  &     16.43 & r &  17.58$^*$ \\
C1-148 & 3C 277.3    & 12 54 11.68 & +27 37 32.70   & II &  $-$4.707 & $-$5.057&  44.959 &  10.892 & H & 1&   $-$2.60 & r &  15.14 \\
C1-155 & 3C 284      & 13 11 08.56 & +27 27 56.50   & II & $-$35.043 & 162.162 &  81.850 &  77.670 & H & 1&     23.20 & r &  17.07 \\
C1-157 & 4C 07.32    & 13 16 20.51 & +07 02 54.30   & I  &           &         &         &         &   &  &     31.79 & r &  13.07 \\
C1-158 & 4C 29.47    & 13 19 06.83 & +29 38 33.80   & I  &    12.449 &   2.925 &   3.498 &   0.517 & L & 1&  $-$17.47 & r &  14.85 \\
C1-162 & 3C 285      & 13 21 21.28 & +42 35 15.20   & II &     9.853 &   4.198 &  37.960 &  15.688 & H & 1&      9.00 & r &  15.65 \\
C1-163 & 4C 03.27    & 13 23 21.04 & +03 08 02.80   & I  &    11.860 &  40.440 &  48.786 & 145.240 & H & 1&   $-$0.51 & r &  16.83 \\
C1-165 & 4C 32.44B   & 13 27 31.71 & +31 51 27.30   & U  &           &         &         &         &   &  &     25.90 & r &  17.17 \\
C1-168 & 3C 287.1    & 13 32 56.37 & +02 00 46.50   & II &    26.772 &  13.199 &  42.994 &  30.812 & H & 1&   $-$3.06 & r &  16.27$^*$ \\
C1-170 & 3C 288      & 13 38 49.67 & +38 51 11.10   & I  &    25.085 &  47.467 &         &         & L & 3&     39.26 & r &  19.46$^*$ \\
C1-172 & 4C 05.57    & 13 42 43.57 & +05 04 31.50   & I  &  $-$2.631 &$-$0.301 &  55.255 &  13.621 & H & 1&  $-$10.81 & r &  15.64 \\
C1-175 & 4C 12.50    & 13 47 33.42 & +12 17 24.10   & C  &    12.737 &  20.474 & 155.231 & 240.504 & H & 1&  $-$11.19 & r &  14.78 \\
C1-176 & 3C 293      & 13 52 17.81 & +31 26 46.70   & I  &    33.077 &  18.370 &  19.589 &   5.356 & L & 1&     29.15 & r &  13.52 \\
C1-185 & S4 1413+34  & 14 16 04.18 & +34 44 36.50   & C  &           &         &         &         &   &  &      9.07 & r &        \\
C1-186 & NGC 5532    & 14 16 53.50 & +10 48 40.20   & I  &    23.832 &   0.543 &  14.906 &   0.253 & L & 1&     37.49 & r &  10.89$^*$ \\
C1-190 & 3C 300      & 14 23 00.81 & +19 35 22.80   & II &           &         &         &         & H & 1&     16.99 & r &  17.92 \\
C1-194 & 4C 07.36    & 14 30 03.34 & +07 15 01.30   & I  &    42.137 &   7.516 &  16.392 &   1.025 & L & 1&  $-$27.35 & r &  13.49 \\
C1-197 & 3C 303      & 14 43 01.45 & +52 01 38.20   & II &     7.362 &   2.848 & 218.611 &  65.857 & H & 1&     24.26 & r &  19.91$^*$ \\
C1-200 & 3C 305      & 14 49 21.74 & +63 16 13.90   & I  &    84.726 &  12.985 &  313.634&  23.957 & H & 1&    106.77 & r &  13.27 \\
C1-203 & B2 1502+28  & 15 04 19.50 & +28 35 34.30   & I  &           &         &         &         &   &  &     56.78 & r &  15.09 \\
      \hline
    \end{tabular}}
\end{table}
\end{landscape}
\begin{landscape}
  \begin{table}
    {\bf Table~\ref{Local} cont.\\}
    \centerline{
    \begin{tabular}{llrrl|rrrrcc|rc|c}
      \hline
      \multicolumn{1}{c}{ID} & \multicolumn{1}{c}{Name} & \multicolumn{1}{c}{RA} & \multicolumn{1}{c}{DEC} & M. & \multicolumn{2}{c}{$\rm [OII]_{3727\AA}$} & \multicolumn{2}{c}{$\rm [OIII]_{5007\AA}$}  & HEG/ & \multicolumn{1}{c}{spec.} & \multicolumn{1}{c}{Richness} & Band & I-mag \\
      & & \multicolumn{2}{c}{(J2000)}    & & \multicolumn{1}{c}{flux} & \multicolumn{1}{c}{EW} & \multicolumn{1}{c}{flux} & \multicolumn{1}{c}{EW} & LEG  & \multicolumn{1}{c}{ref}& & & \\
        \hline 
C1-205 & 3C 310      & 15 04 58.98 & +25 59 49.00   & I  &    32.062 &  22.572 &  11.807 &   2.749 & L & 1 &     39.47 & r &  14.71 \\
C1-209 & 3C 315      & 15 13 39.90 & +26 07 33.70   & I  &           &         &         &         & L & 1 &   $-$7.66 & r &  16.30 \\
C1-211 & 4C 00.56    & 15 16 40.21 & +00 15 02.40   & II &    30.807 &   5.221 &  78.387 &  19.001 & H & 1 &     49.18 & r &  14.17 \\
C1-216 & 3C 319      & 15 24 05.64 & +54 28 18.40   & II &           &         &         &         & L & 1 &     23.24 & r &  17.67 \\
C1-219 & 3C 321      & 15 31 50.71 & +24 02 43.30   & II &           &         &         &         & H & 1 &   $-$6.63 & r &  15.35 \\
C1-226 & 3C 323.1    & 15 47 44.23 & +20 52 41.00   & II &    12.246 &   0.106 & 283.969 &  26.345 & H & 1 &     17.79 & r &  15.11 \\
C1-230 & 3C 326      & 15 52 26.86 & +20 05 01.80   & II &           &         &         &         & L & 1 &     39.58 & r &  15.99 \\
C1-234 & 3C 327      & 16 02 17.21 & +01 58 19.40   & II &           &         &         &         & H & 1 &      2.88 & R &  13.92$^*$ \\
C1-242 & NGC 6109    & 16 17 38.89 & +35 00 48.00   & I  &    23.121 &   2.414 &  12.021 &   0.120 & L & 1 &      3.51 & r &  12.77 \\
C1-243 & 3C 332      & 16 17 43.28 & +32 23 02.40   & II &    46.467 &   5.868 & 179.076 &  46.312 & H & 1 &   $-$8.99 & r &  16.06 \\
C1-248 & 3C 338      & 16 28 38.34 & +39 33 04.70   & I  &           &         &         &         & L & 1 &     78.57 & r &   6.16 \\
C1-258 & 3C 346      & 16 43 48.69 & +17 15 48.80   & I  &           &         &         &         & H & 1 &     44.37 & r &  15.92 \\
C1-260 & 4C 39.49    & 16 53 52.24 & +39 45 36.60   & C  &           &         &         &         &   &   &     11.22 & r &  12.97 \\
C1-261 & 3C 349      & 16 59 27.57 & +47 03 13.10   & II &           &         &         &         & H & 1 &     27.80 & r &  19.42$^*$ \\
C1-266 & 4C 34.47    & 17 23 20.85 & +34 17 57.30   & II &           &         &         &         & H &5  &     15.37 & r &  14.94 \\
C1-270 & 3C 306      & 14 54 20.30 & +16 20 55.80   & II &     9.610 &   0.703 &$-$2.929 &$-$1.692 & L & 1 &     33.59 & r &  12.79 \\
C1-271 & 4C 32.25A   & 08 31 20.33 & +32 18 37.00   & II &    13.967 &   3.779 &  16.312 &   3.007 & L & 1 &      7.28 & r &  14.14 \\
C1-272 & 4C 06.32    & 08 48 41.94 & +05 55 35.00   & II &           &         &         &         &   &   &     63.73 & r &  17.31 \\
C2-031 & 4C 21.26    & 09 54  7.03 & +21 22 35.90   & II &    12.518 &   4.216 & 150.113 &  76.453 & H & 1 &   $-$3.36 & r &  16.61 \\
C2-041 & 4C 20.20    & 10 02 57.12 & +19 51 53.50   & I  &  $-$1.185 &$-$0.339 &   1.464 &   0.515 & L & 1 &     27.49 & r &  16.39 \\
C2-045 & 4C 13.41    & 10 07 26.10 & +12 48 56.21   & II & $-$11.080 &$-$2.362 & 115.009 &   3.945 & H & 1 &     16.52 & r &  14.06 \\
C2-049 & 4C 14.36    & 10 09 55.50 & +14 01 54.10   & C  &     4.908 &  10.551 &   4.382 &   2.530 & L & 1 &     10.47 & r &  16.71 \\
C2-055 & 4C 41.22    & 10 15 58.26 & +40 46 47.11   & II &     2.190 &   0.901 &   1.058 &   0.233 & L & 1 &     10.24 & r &  15.81 \\
C2-067 & 3C 244      & 10 27 32.89 & +48 17  6.40   & II &  $-$2.615 &  15.726 &  18.087 &  35.759 & H & 1 &     23.67 & r &  17.85 \\
C2-070 & 4C 52.22    & 10 31 43.55 & +52 25 37.90   & II &     4.366 &   5.374 &  20.252 &  13.394 & H & 1 &     16.58 & r &  16.67 \\
C2-102 & 1108+201    & 11 11 20.09 & +19 55 36.10   & C  &    20.200 &         &  10.700 &         & H &3  &     28.37 & r &  17.54 \\
C2-105 & 4C 41.23    & 11 11 43.62 & +40 49 15.30   & I  &     4.393 &   0.567 &   1.322 &   0.062 & L & 1 &     58.80 & r &  14.17 \\
C2-117 & 4C 05.50    & 11 24 37.45 & +04 56 18.80   & II &    10.810 &  76.245 &  64.728 & 231.621 & H & 1 &  $-$12.06 & r &  16.83 \\
C2-118 & 3C 258      & 11 24 43.90 & +19 19 29.70   & C  &           &         &         &         & L &3  &  $-$11.53 & r &  17.95 \\
C2-123 & 4C 00.40    & 11 29 35.97 & +00 15 17.50   & II &           &         &         &         &   &   &     30.71 & r &  17.52$^*$ \\
C2-127 & 4C 33.27    & 11 33  9.56 & +33 43 12.60   & II & $-$49.036 &1440.820 &   3.835 &   2.388 & L & 1 &  $-$13.68 & r &  16.73 \\
C2-134 & 4C 17.52    & 11 40 17.03 & +17 43 39.00   & I  &           &         &  13.402 &   0.796 & L & 1 &      9.06 & r &  17.26$^*$ \\
C2-141 & 4C 46.23    & 11 43 39.63 & +46 21 20.70   & II &     9.633 &  12.989 &   4.282 &   1.999 & L & 1 &     56.20 & r &  15.79 \\
C2-162 & 1155+251    & 11 58 25.80 & +24 50 17.70   & C  &    10.786 &  26.324 &  47.062 &  91.354 & H & 1 &   $-$9.18 & r &  16.93 \\
C2-169 & 4C 58.23    & 12 02  4.19 & +58 02  1.90   & I  &     2.105 &   0.428 &$-$0.982 &$-$0.910 & L & 1 &     67.83 & r &  15.67$^*$ \\
C2-200 & 1227+181    & 12 29 32.62 & +17 50 20.90   & C  &           &         &         &         &   &   &     22.16 & r &  16.93 \\
C2-214 & 4C 49.25    & 12 47  7.40 & +49 00 18.20   & C  &     6.053 &  10.609 &  10.304 &   9.312 & H & 1 &   $-$5.85 & r &  17.04 \\
C2-220 & 1249+035    & 12 52 22.78 & +03 15 50.40   & I  &     4.271 &   1.144 &$-$1.819 &   0.002 & L & 1 &  $-$19.98 & r &  14.68 \\
C2-226 & 4C 44.22    & 12 58  1.96 & +44 35 20.60   & II &           &         &         &         &   &   &     30.32 & r &  17.07 \\
C2-239 & 4C 08.38    & 13 15  9.94 & +08 41 44.60   & II &    12.186 &   7.178 &  20.918 &   4.739 & L & 1 &     38.92 & r &  16.13 \\
C3-007 & 1440+163    & 14 43  1.74 & +16 06 59.90   & II &           &         &         &         &   &   &     19.81 & r &  17.13 \\
C3-010 & 1441+25     & 14 43 56.94 & +25 01 44.50   & C  &           &         &         &         &   &   &  $-$23.87 & r &  18.60 \\
C3-015 & B1442+195   & 14 44 34.84 & +19 21 33.00   & I  &     0.563 &   0.076 &$-$1.572 &$-$0.067 & L & 1 &     35.81 & r &  16.50 \\
C3-021 & 4C 17.60    & 14 45 57.34 & +17 38 30.20   & II &     8.717 &   1.066 &   4.359 &   0.183 & L & 1 &      1.49 & r &  15.33 \\
C3-029 & 4C 16.42    & 14 47 44.55 & +16 36  6.00   & II &           &         &         &         &   &   &  $-$12.32 & r &  19.52$^*$ \\
C3-030 & 1445+149    & 14 48  4.28 & +14 47  4.60   & I  &  $-$2.085 &  16.911 &$-$0.667 &$-$2.205 & L & 1 &     65.85 & r &  16.27 \\
C3-032 & 1446+277    & 14 48 27.87 & +27 33 18.80   & C  &     0.683 &   2.128 &$-$0.600 &$-$0.769 & L & 1 &     47.77 & r &  17.34 \\
C3-034 & 3C 304      & 14 48 50.05 & +20 25 34.80   & II &           &         &         &         &   &   &      5.64 & r &  17.99 \\
      \hline
    \end{tabular}}
\end{table}
\end{landscape}
\begin{landscape}
  \begin{table}
    {\bf Table~\ref{Local} cont.\\}
    \centerline{
    \begin{tabular}{llrrl|rrrrcc|rc|c}
      \hline
      \multicolumn{1}{c}{ID} & \multicolumn{1}{c}{Name} & \multicolumn{1}{c}{RA} & \multicolumn{1}{c}{DEC} & M. & \multicolumn{2}{c}{$\rm [OII]_{3727\AA}$} & \multicolumn{2}{c}{$\rm [OIII]_{5007\AA}$}  & HEG/ & \multicolumn{1}{c}{spec.} & \multicolumn{1}{c}{Richness} & Band & I-mag \\
      & & \multicolumn{2}{c}{(J2000)}    & & \multicolumn{1}{c}{flux} & \multicolumn{1}{c}{EW} & \multicolumn{1}{c}{flux} & \multicolumn{1}{c}{EW} & LEG  & \multicolumn{1}{c}{ref}& & & \\
        \hline 
C3-035 & 1447+213    & 14 49 19.01 & +21 05 48.00   & II &           &         &         &         &   &   &      8.75 & r &   17.34 \\
C3-052 & 1452+258    & 14 54 22.75 & +25 39 55.50   & II &           &         &         &         &   &   &      0.07 & r &   18.57$^*$ \\
C3-056 & 1452+144    & 14 55  7.32 & +14 12 22.20   & II &           &         &         &         &   &   &      4.73 & r &   19.40$^*$ \\
C3-057 & NGC 5782    & 14 55 55.36 & +11 51 44.70   & I  &     5.401 &   0.095 &$-$4.513 &$-$0.138 & L & 1 &    103.60 & r &   12.67 \\
C3-058 & 4C 16.43    & 14 56  5.65 & +16 26 52.80   & II &  $-$1.515 &   7.108 &   1.387 &   2.462 & L & 1 &     12.13 & r &   17.24 \\
C3-069 & 4C 28.38    & 14 57 53.80 & +28 32 20.00   & II &     3.215 &   3.622 & 174.347 &  76.256 & H & 1 &     10.94 & r &   16.32 \\
C3-078 & B2 1457+29  & 14 59 42.07 & +29 03 34.10   & II &           &         &         &         &   &   &     14.08 & r &   16.51 \\
C3-079 & 1458+204    & 15 00 24.05 & +20 12 37.80   & I  &           &         &         &         &   &   &      8.60 & r &   14.84 \\
C3-080 & 4C 14.57    & 15 00 21.36 & +14 34 59.80   & II &           &         &         &         &   &   &  $-$16.15 & r &   15.48 \\
C3-082 & 4C 21.44    & 15 01 28.50 & +21 34 20.70   & I  &     1.103 &   1.946 &   5.545 &   0.596 & L & 1 &     59.30 & r &   16.61 \\
C3-089 & 1500+1832   & 15 03  1.63 & +18 20 32.40   & II &           &         &         &         &   &   &     11.88 & r &   18.36 \\
C3-093 & MRC1501+104 & 15 03 39.51 & +10 16  2.80   & I  &           &         &  19.000 &         & H & 1 &     15.71 & r &   15.92 \\
C3-104 & B1507+105   & 15 07 21.88 & +10 18 46.30   & C  &    59.922 &  15.452 &   9.584 &   1.110 & L & 1 &   $-$5.00 & r &   14.07 \\
C3-117 & J1509+1557  & 15 09 50.53 & +15 57 25.70   & C  &     6.182 &   6.974 &   9.215 &   3.985 & L & 1 &     15.11 & r &   16.32 \\
C3-125 & 1508+182    & 15 11  9.08 & +18 01 53.80   & I  &     1.246 &   0.502 &   0.042 &$-$0.409 & L & 1 &     31.78 & r &   15.42 \\
C3-137 & 1511+2422   & 15 13 45.74 & +24 11  2.80   & II &     6.207 &   3.272 &   5.773 &   1.588 & L & 1 &     12.60 & r &   15.39 \\
C3-138 & 1511+225    & 15 14  3.55 & +22 23 31.50   & C  &           &         &         &         &   &   &      2.04 & r &   16.96 \\
C3-139 & 1512+2338   & 15 14 14.64 & +23 27 11.20   & II &     0.741 &   0.094 &$-$3.213 &$-$1.644 & L & 1 &   $-$2.88 & r &   15.18 \\
C3-142 & 1512+104    & 15 14 49.50 & +10 17  0.90   & I  &  $-$3.461 &$-$2.461 &$-$5.581 &$-$1.513 & L & 1 &   $-$9.06 & r &   14.06 \\
C3-146 & 1513+144    & 15 16  2.98 & +14 18 22.90   & II &           &         &         &         &   &   &  $-$28.34 & r &   17.65 \\
C3-149 & 1514+215    & 15 17  4.56 & +21 22 42.90   & II &           &         &         &         &   &   &      3.92 & r &   18.13 \\
C3-151 & 1515+176    & 15 17 24.70 & +17 29 28.30   & II &     5.942 &  10.378 &  79.524 &  98.106 & H & 1 &     48.94 & r &   17.65 \\
C3-165 & 1519+153    & 15 21 16.47 & +15 12  9.90   & U  &     9.338 &  37.183 &   6.619 &   7.706 & L & 1 &     21.39 & r &   16.82 \\
C3-166 & 1519+108    & 15 22 12.15 & +10 41 31.00   & II &           &         &         &         &   &   &     27.23 & r &   17.22 \\
C3-167 & 1519+103    & 15 22 17.09 & +10 13  0.50   & II &           &         &         &         &   &   &   $-$4.91 & r &   18.11 \\
C3-172 & 1521+116    & 15 23 27.56 & +11 30 23.90   & I  &     4.051 &   4.476 &   6.282 &   9.386 & H & 1 &   $-$8.30 & r &   16.87 \\
C3-173 & 4C 28.39    & 15 23 28.40 & +28 36  4.10   & I  &     5.511 &   1.056 &   1.083 &   1.006 & L & 1 &      9.11 & r &   14.87 \\
C3-181 & 1522+130    & 15 25  8.80 & +12 53 18.10   & II &     2.033 &   4.946 &  27.160 &  34.274 & H & 1 &      2.11 & r &   18.05 \\
C3-189 & 1525+290    & 15 27 44.61 & +28 55  6.60   & I  &           &         &         &         &   &   &     35.94 & r &   14.65 \\
C3-190 & 1525+227    & 15 27 57.80 & +22 33  1.30   & II &     8.748 &   0.323 & 107.179 &  14.085 & H & 1 &   $-$3.95 & r &   16.42 \\
C3-195 & 1528+29     & 15 30  4.69 & +29 00  9.30   & II &           &         &         &         &   &   &  $-$35.12 & r &   14.94 \\
C3-196 & 1527+234    & 15 30  5.11 & +23 16 22.20   & II &     0.925 &   0.071 &$-$3.081 &$-$1.432 & L & 1 &     41.16 & r &   15.19 \\
C3-203 & B2 1530+28  & 15 32 44.30 & +28 03 46.40   & I  &     2.405 &   0.435 &$-$2.623 &$-$1.453 & L & 1 &     71.95 & r &   16.37 \\
C3-208 & 1531+104    & 15 34 17.83 & +10 17  8.40   & I  &           &         &         &         &   &   &     55.93 & r &   16.28 \\
C3-209 & 1532+139    & 15 34 22.66 & +13 49 17.10   & II &           &         &         &         &   &   &     18.42 & r &   17.22 \\
C3-211 & ARP 220     & 15 34 57.26 & +23 30 11.10   & C  &           &         &  14.129 &   7.417 & H & 1 &  $-$15.41 & r &   13.23 \\
C3-216 & 1534+269    & 15 37  7.76 & +26 48 28.50   & I  &           &         &         &         &   &   &     16.65 & r &   17.66 \\
C3-231 & 1541+230    & 15 43 28.53 & +22 52 32.80   & II &     0.190 &   0.371 &   1.368 &   0.178 & L & 1 &     29.22 & r &   15.90$^*$ \\
C3-244 & 1545+1505   & 15 47 30.07 & +14 56 55.70   & I  &     6.530 &   1.661 &   1.756 &$-$0.103 & L & 1 &      6.87 & r &   15.07 \\
C3-266 & 4C 23.42    & 15 53 43.61 & +23 48  4.70   & I  &    33.605 &  12.983 &  42.710 &   7.505 & H & 1 &      4.65 & r &   15.36 \\
C3-282 & 4C 10.44    & 15 56 47.07 & +10 37 55.70   & I  &     0.734 &   0.758 &   0.913 &   0.202 & L & 1 &     47.93 & r &   16.14 \\
C3-284 & 4C 12.56    & 15 59  6.89 & +12 10 26.90   & II &           &         &         &         &   &   &      2.36 & r &   18.08 \\
C4-002 & 1405+026    & 14 08 28.14 & +02 25 48.70   & I  &           &         &         &         &   &   &      0.62 & r &   17.76 \\
C4-014 & 1409-0307   & 14 09 52.02 & $-$03 03 10.30 & II &     2.693 &   2.179 &   2.361 &   0.684 & L & 1 &     54.38 & r &   15.76 \\
C4-016 & 1409-0135   & 14 09 57.00 & $-$01 21  4.70 & I  &           &         &         &         &   &   &     29.73 & r &   18.12 \\
C4-028 & 1411+0229   & 14 11 14.61 & +02 17 22.50   & U  &     3.005 &   0.530 &$-$0.444 &$-$2.266 & L & 1 &      6.68 & r &   17.94 \\
C4-036 & NGC 5506    & 14 13 14.84 & $-$03 12 27.00 & I  &           &         &3335.108 & 539.967 & H & 1 &  $-$27.96 & r &   10.84$^*$ \\ 
C4-044 & 1414+0182   & 14 14  9.37 & +01 49 10.80   & II &  $-$0.404 &$-$1.918 &$-$1.267 &$-$0.164 & L & 1 &     12.51 & r &   16.67 \\
      \hline
    \end{tabular}}
\end{table}
\end{landscape}
\begin{landscape}
  \begin{table}
    {\bf Table~\ref{Local} cont.\\}
    \centerline{
    \begin{tabular}{llrrl|rrrrcc|rc|c}
      \hline
      \multicolumn{1}{c}{ID} & \multicolumn{1}{c}{Name} & \multicolumn{1}{c}{RA} & \multicolumn{1}{c}{DEC} & M. & \multicolumn{2}{c}{$\rm [OII]_{3727\AA}$} & \multicolumn{2}{c}{$\rm [OIII]_{5007\AA}$}  & HEG/ & \multicolumn{1}{c}{spec.} & \multicolumn{1}{c}{Richness} & Band & I-mag \\
      & & \multicolumn{2}{c}{(J2000)}    & & \multicolumn{1}{c}{flux} & \multicolumn{1}{c}{EW} & \multicolumn{1}{c}{flux} & \multicolumn{1}{c}{EW} & LEG  & \multicolumn{1}{c}{ref}& & & \\ 
        \hline 
C4-047 & LEDA 184576 & 14 14 57.34 & +00 12 17.90   & I  &          &         &         &          &   & 6 &   15.74 & r &  17.81 \\
C4-049 & N274Z243    & 14 15 11.41 & $-$01 37  2.80 & I  &     0.376& $-$0.031&   0.494 &    0.156 & L & 1 &   26.57 & r &  15.35 \\
C4-050 & N342Z086    & 14 15 28.72 & +01 05 54.20   & I  &          &         &         &          & H & 6 &    3.58 & r &  16.12 \\
C4-055 & 1416+0219   & 14 16 13.74 & +02 19 22.50   & I  &     4.336&    2.100& 163.591 &   58.062 & H & 1 &   22.83 & r &  15.92 \\
C4-056 & J141643-02  & 14 16 43.04 & $-$02 56 11.30 & C  &     0.990&    3.815&   2.372 &    1.533 & L & 1 & $-$5.64 & r &  16.63 \\
C4-085 & N344Z154    & 14 24  3.40 & +00 29 58.70   & I  &     0.199&    0.025&$-$1.419 & $-$0.764 & L & 1 & $-$1.49 & r &  15.09 \\
C4-098 & N344Z014    & 14 26 15.51 & +00 50 21.70   & I  &  $-$0.460&    0.664&$-$1.119 & $-$1.316 & L & 1 &   15.88 & r &  15.42 \\
C4-143 & 1433-0239   & 14 33 46.69 & $-$02 23 22.50 & I  &          &         &         &          &   &   &$-$19.96 & r &  17.34 \\
C4-146 & 1434+0158   & 14 34 10.56 & +01 36 46.90   & I  &     0.701& $-$0.213&$-$2.508 & $-$1.444 & L & 1 &   37.21 & r &  15.29 \\
C4-150 & 1432-020    & 14 34 49.27 & $-$02 15  9.20 & II &     2.694&    3.408&   0.820 &    0.091 & L & 1 &   39.54 & r &  17.53 \\
C4-155 & 1436+0181   & 14 36  9.04 & +01 48 49.20   & C  &          &         &         &          &   &   &   29.71 & r &  19.86$^*$ \\
C4-166 & 1437-0025   & 14 37 42.80 & $-$00 15  4.20 & I  &     0.737&    0.266&$-$3.489 & $-$0.288 & L & 1 &   61.25 & r &  15.50 \\
C4-176 & 1438-0133   & 14 38 20.57 & $-$01 20  6.60 & II &          &         &         &          &   &   &$-$13.49 & r &  17.53 \\
C4-178 & 1438-0100   & 14 38 25.93 & $-$01 00  1.50 & I  &          &         &         &          & L & 6 &   30.49 & r &  19.65$^*$ \\
C4-184 & 1438+0068   & 14 38 48.87 & +00 40 59.20   & I  &    18.829&   29.574&  18.254 &   16.191 & H & 7 &    7.21 & r &  16.14$^*$ \\
CE-008 & CE-008      & 09 57 30.07 & $-$21 30 59.80 & II &          &         &         &          & L & 7 &   48.94 & I &  17.77 \\
CE-009 & CE-009      & 09 49 35.43 & $-$21 56 23.50 & C  &          &         &         &          & H & 7 &   21.24 & I &  18.28 \\
CE-018 & CE-018      & 09 55 13.60 & $-$21 23  3.10 & C  &          &         &         &          & H & 7 &   40.64 & I &  14.88 \\
CE-030 & CE-030      & 09 45 55.86 & $-$20 28 30.20 & I  &          &         &         &          & L & 7 &   99.51 & I &  16.41 \\
CE-041 & CE-041      & 09 49 18.18 & $-$20 54 45.40 & I  &          &         &         &          & L & 7 &    8.65 & I &  17.10 \\
CE-075 & CE-075      & 09 45 26.97 & $-$20 33 55.00 & II &          &         &         &          & L & 7 &   77.71 & I &  16.73 \\
CE-076 & CE-076      & 09 57 45.89 & $-$21 23 23.60 & C  &          &         &         &          & L & 7 &   25.05 & I &  17.23 \\
CE-084 & CE-084      & 09 55 45.19 & $-$21 25 23.00 & II &          &         &         &          & H & 7 & $-$0.91 & I &  15.11 \\
CE-093 & CE-093      & 09 46 18.86 & $-$20 37 57.40 & I  &          &         &         &          & L & 7 &   10.25 & I &  17.41 \\
CE-095 & CE-095      & 09 54 21.48 & $-$21 48  7.20 & U  &          &         &         &          & H & 7 &  153.05 & I &  16.92 \\
CE-108 & CE-108      & 09 56 49.76 & $-$20 35 25.90 & C  &          &         &         &          & L & 7 &$-$13.43 & I &  17.09 \\
CE-110 & CE-110      & 09 55 11.49 & $-$20 30 18.70 & I  &          &         &         &          & L & 7 &   20.82 & I &  17.52 \\
CE-121 & CE-121      & 09 52  1.20 & $-$20 24 56.50 & C  &          &         &         &          & H & 7 &$-$17.36 & I &  17.15 \\
CE-122 & CE-122      & 09 56 37.11 & $-$20 19  5.50 & II &          &         &         &          & L & 7 &   22.47 & I &  16.94 \\
      \hline
    \end{tabular}}
\end{table}
\end{landscape}

\label{lastpage}

\end{document}